\documentclass[12pt,preprint]{aastex}

\newcommand{\msun} {$M_{\sun}$}

\newcommand{\Te} {T_{\rm eff}}
\newcommand{\logg} {\log g}

\begin{document}

\title{AN IMPROVED SPECTROSCOPIC ANALYSIS OF DA WHITE DWARFS 
FROM THE SLOAN DIGITAL SKY SURVEY DATA RELEASE 4}

\author{P.-E. Tremblay, P. Bergeron, and A. Gianninas}
\affil{D\'epartement de Physique, Universit\'e de Montr\'eal, C.P.~6128, 
Succ.~Centre-Ville, Montr\'eal, Qu\'ebec H3C 3J7, Canada.}
\email{tremblay@astro.umontreal.ca, bergeron@astro.umontreal.ca, gianninas@astro.umontreal.ca}

\begin{abstract}

We present an improved spectroscopic and photometric analysis of
hydrogen-line DA white dwarfs from the Sloan Digital Sky Survey Data
Release 4 based on model atmospheres that include improved Stark
broadening profiles with non-ideal gas effects. We also perform a
careful visual inspection of all spectroscopic fits with high
signal-to-noise ratios (S/N $>$ 12) and present improved atmospheric
parameters ($\Te$ and $\log g$) for each white dwarf. Through a
comparison of spectroscopic and photometric temperatures, we report
the discovery of 35 DA+DB/DC double degenerate candidates and 2
helium-rich DA stars.  We also determine that a cutoff at S/N = 15
optimizes the size and quality of the sample for computing the mean
mass of DA white dwarfs, for which we report a value of 0.613
$M_{\odot}$. In the following step, we compare our results to previous
analyses of the SDSS DR4 and find a good agreement if we account for
the shift produced by the improved Stark profiles. Finally, the
properties of DA white dwarfs in the SDSS are weighed against those of
the Villanova White Dwarf Catalog sample of Gianninas et al. We find
systematically lower masses (by about $3\%$ on average), a difference
that we trace back to the data reduction procedure of the SDSS. We
conclude that a better understanding of these differences will be
important to determine the absolute temperature scale and mean mass of
DA white dwarfs.

\end{abstract}


\section{INTRODUCTION}

The now completed Sloan Digital Sky Survey \citep[SDSS;][]{sdss} has
rapidly grown to be, by far, the largest source of newly discovered
white dwarfs in our galaxy. This project started as a survey of faint
blue point sources and finally covered 9380 square degrees of the sky
in the latest data release \citep[][Data Release 7]{dr7}. They mostly
surveyed the North Galactic Cap region, with a completeness of about
95$\%$ for the point-source $ugriz$ photometry. They performed a
spectroscopic follow-up of nearly 1.6 million objects, mainly selected
for their blue colors. White dwarfs from the
different data releases have been identified in a series of papers
\citep{harris03,kleinman04,harris06}, with the latest catalog
published by \citet{E06} containing 9316 white dwarfs drawn from the
Data Release 4 (the SDSS-E06 catalog hereafter), covering 4783 square
degrees of the sky. The final catalog from the DR7 is currently in
preparation \citep{kleinman09} and will increase the total number of
white dwarfs discovered in the SDSS by nearly a factor of two. While
the SDSS survey has been the source of numerous discoveries of
peculiar objects (this subject is too wide to review here), and even a
new class of white dwarfs --- the so called hot-DQ stars
\citep{dufour07}, the most common class of objects -- the
hydrogen-line (DA) white dwarfs -- has been at the center of only a
few studies \citep{E06,kepler07}.

The largest published sample of spectroscopically identified DA white
dwarfs can be found in the SDSS-E06 catalog. Considering the large
number of SDSS spectra of all types, only the {\it most obvious} white
dwarfs were recovered in this catalog. The main problem in the
identification of white dwarfs in the SDSS is that at cool
temperatures ($\Te\lesssim 8000$~K), they overlap in a color-color
diagram with A and F main-sequence stars. However, for hotter objects,
they claim their catalog should recover most of the single DA stars
observed spectroscopically by the SDSS. We will review their selection
procedure in more detail in Section 2. The catalog provided individual
atmospheric parameters --- $\Te$ and $\logg$ --- for each DA white
dwarf, but this was not meant to be a careful analysis of these stars,
especially given that many objects have been catalogued without a
visual inspection. The only thorough follow-up analysis, so far, of
the bulk of these white dwarfs has been reported by \citet{kepler07},
who refitted more carefully all objects with $g<19$, and concentrated
in particular on the mass distribution, including a comparison with
previous determinations from independent surveys, and a comparison of
the mass distribution of DA stars with that obtained for the
helium-line (DB) white dwarfs identified in the SDSS-E06. However,
Kepler et al.~did not provide any update of the atmospheric parameters
(which differed from those published by Eisenstein et al.).
\citet{degennaro08} reviewed the luminosity function and
completeness of the SDSS-E06 sample, but used the atmospheric
parameters previously published by Eisenstein et al. In parallel,
subclasses of the DA spectral type --- e.g. magnetic white dwarfs, DAO
stars with detectable He~\textsc{ii} $\lambda4686$, and the DA$-$M
dwarf binaries --- have been analyzed in various independent studies,
covering partially, or in full, the SDSS-E06 catalog.

We have recently computed a new set of model atmospheres for
DA white dwarfs, described in Section 2.2, which rely on the improved
Stark broadening profiles developed by \citet{TB09}.  These model
spectra have already been applied to the hot white dwarfs in the
Palomar-Green (PG) sample \citep{TB09}, and they are also currently being
applied to the large scale analysis of the DA stars in the Villanova
White Dwarf Catalog\footnote{http://www.astronomy.villanova.edu/WDCatalog/index.html}
\citep{gianninas09,gianninas11}.  
In this paper, the same models are used in an independent analysis of
the hydrogen-line (DA) stars identified in the SDSS-E06 catalog.  In
particular, we will obtain improved stellar parameters, which can be
readily compared to those found in the PG sample and the Gianninas et
al.~sample based on the same model spectra and fitting technique. This
will allow for a more robust test of the quality of the SDSS data
reduction and sample selection.

The SDSS sample is unique in the sense that it includes many more
objects than any other sample of white dwarfs previously
compiled. This is certainly an advantage since it provides better
statistics on rare peculiar objects. However, for the bulk of the
regular DA stars, we can see in Figure \ref{fg:f1} that the average
signal-to-noise ratio (S/N) of the observed spectra is low, much lower
than any other sample previously analyzed \citep[see, e.g., Fig.~1
  of][]{LBH05}. Therefore, we think it is essential to better quantify
the reliability of lower S/N data, a point also raised by
\citet{koester09}. \citet{kepler07} used a simple magnitude cutoff,
but here we propose a more detailed study of how the size of the
sample can be optimized to yield average properties of the sample that
have the most statistical significance. Finally, we have carried out a
careful visual inspection of all the fits for objects with ${\rm S/N} >
12$, a task that has not been accomplished until now, to our
knowledge. In the course of this inspection, we have identified many
new peculiar objects, misclassified in previous analyses, including a
large number of degenerate binaries.

The goals outlined in the previous paragraphs led us to perform a
complete and detailed follow-up of the studies of Eisenstein et
al.~and Kepler et al.~by re-analyzing all DA white dwarfs in the
SDSS-E06 sample. In Section 2, we summarize the set of SDSS
observations and also describe our model atmospheres and fitting
technique. We then present the results of our updated analysis in
Section 3, where we define our optimal sample and also report the
discovery of new degenerate binaries. Our results are compared in
Section 4 with those of previous analyses of the SDSS-E06 sample as
well as other samples of DA stars. Concluding remarks follow in
Section 5.

\section{DATA ANALYSIS}

\subsection{The SDSS Sample of DA White Dwarfs}

The SDSS-E06 catalog relies on an automatic procedure to recover the
white dwarf spectra in the survey. We briefly summarize this procedure
in three main steps. A blue portion of the $g-r$ vs.~$u-g$ diagram
(see Fig.~1 of \citealt{E06}) is first defined to identify white dwarf
candidates. As mentioned in the Introduction, white dwarfs cooler than
$\sim$8000~K occupy the same region as main-sequence stars in this
two-color diagram, and most of them are therefore excluded by this
first color cutoff. A second step is to eliminate candidates with a
galaxy classification and a redshift larger than 0.003, although a few
of them remain when a proper motion measurement is available with a
value consistent with a nearby galactic source. Finally, the resulting
photometric and spectroscopic observations for $\sim$13,000 white
dwarf candidates are compared to white dwarf models with a $\chi^2$
minimization fitting technique. The outliers and objects with a poor
fit are reclassified manually and in some cases rejected from the
final catalog altogether. These steps seem quite robust in recovering
all typical, hot, single DA white dwarfs. The authors of the catalog
claim that it should be fairly complete for DA white dwarfs with $\Te
\gtrsim 8000$~K. However, they also remind us that the completeness of
the SDSS spectroscopic survey itself can be anywhere between 15\% and
50\%, in part due to the neglect of blended point sources, as well as
an incomplete spectroscopic follow-up of the point sources identified
in the SDSS fields \citep{E06,degennaro08}. For all other DA subtypes
(e.g, magnetics, DA$-$M dwarf binaries, etc.), the picture is less
clear. The most common subtype are the DA -- M dwarf binaries where
the presence of a cool main-sequence companion often changes the
observed colors and contaminates the line profiles. The SDSS-E06
catalog is not tuned to identify systematically these objects. Other
subtypes with spectra significantly different from that of a normal DA
white dwarf, such as magnetics or DA stars diluted by a featureless DC
companion, are also more likely to be rejected.

The basis of our analysis are the 8717 spectroscopic observations from
the SDSS-E06 catalog with spectral type DA as the main classification,
including some multiple spectra of the same objects. We obtained the
spectroscopic and photometric observations from the SDSS Data Archive
and Sky Server\footnote{das.sdss.org, cas.sdss.org}. The spectra,
which cover a wavelength range of 3800 to $\sim$10,000 \AA~with a
resolution of $R\sim 1800$, rely on the improved data reduction from
the Data Release 7, where some data reduction problems observed in
previous data releases have been corrected \citep{kleinman09}. We next
computed the S/N for each object in the range from 4450 to 4750~\AA, a
featureless spectral region between H$\beta$ and H$\gamma$, which
provides a representative average for our spectroscopic technique. As
expected, we can see in Figure \ref{fg:f1} that the S/N of the spectra
is closely correlated with the observed $g$ magnitudes. We prefer to
constrain the size of our sample based on S/N rather than magnitude
because the former value is more directly related to the uncertainties
in our fitting procedure described below. It is immediately clear that
the faintest objects in the SDSS will not be included in the mean
properties computation, because their atmospheric parameters carry
large individual uncertainties. It also becomes increasingly difficult
to identify DA subtypes (e.g. magnetics, DAB) when the noise level is
high. For these reasons, we break the complete sample of DA spectra
into two categories. The higher quality data (${\rm S/N}>12$), for a
total of 3249 stars (3550 different observations), are carefully
analyzed through a visual inspection of each fit. The remaining lower
quality data of the sample are fitted with automatic programs and used
for reference only. Unless it is noted explicitly, all the numbers and
figures in our analysis are based on the higher quality data only. The
reason for the ${\rm S/N}=12$ cutoff is simply that a visual
inspection of the lower quality spectra would not yield a significant
number of interesting detections, and the publication of updated
atmospheric parameters was deemed of low importance.

\subsection{Model Atmospheres}

For this work, we use the improved DA white dwarf model spectra that
have been developed by our group in recent years. First of all, we
rely on the Stark broadening tables of \citet{TB09} with the non-ideal
gas effects from \citet{HM88} included directly in the line profile
calculations. Our models are similar to those used in their analysis
in the range 40,000~K $> \Te >$ 12,000~K, with the exception that we
now add the non-ideal effects due to the electronic collisions in the
equation of state as well (i.e., in addition to the line profiles). \citet[][see
Sec.~2.3]{TB09} argued that these effects could be neglected compared
to proton perturbations, but we actually find that this physical
ingredient still changes the mean properties by $\sim$1\%. For cooler
temperatures, we employ the same model grid as that described in
\citet{TB10}. In particular, we make use of the ML2/$\alpha=0.8$
version of the mixing-length theory, which provides, within the
context of our improved broadening profiles, the best overall internal
consistency between optical and UV temperatures.

For the purpose of analyzing the hottest DA stars in the SDSS sample,
we also computed new non-LTE model spectra with $\Te > 40,000$~K using
the publically available TLUSTY model atmosphere code and the
accompanying spectral synthesis extension SYNSPEC \citep{TLUSTY}. The
only difference between this grid and that previously described in
\citet{LBH05} is that we are now using our improved Stark broadening
profiles, and the $\beta_{\rm crit}$ parameter was also set back to
its original value (see \citealt{TB09} for details). It is worth
mentioning that we include these new Stark profiles in both the
atmospheric structure (TLUSTY) and model spectrum (SYNSPEC)
calculations, while previous non-LTE grids calculated with TLUSTY
relied on approximate analytical Stark profiles, based on the
two-level approximation (similar to the assumptions of
\citealt{lemke97}), to compute the atmospheric structures. Because
such analytical profiles could not be easily modified to take into
account non-ideal effects, it was deemed necessary to replace these
approximate profiles with our more detailed calculations. This
improvement ensures that our Stark profiles are now included in a
self-consistent way throughout all our calculations. Our complete grid
of model atmospheres and spectra covers a range of $\Te=1500$~K to
140,000 K (by steps of 250 K for $\Te<5500$~K, 500 K for 6000~K $<$
$\Te$ $<$ 17,000~K, 5000~K for 20,000~K $<$ $\Te$ $<$ 90,000~K, and
10,000~K for $\Te$ $>$ 90,000~K), with log g values from 6.0 to 10.0 (by
steps of 0.5 dex with additional points at 7.75 and 8.25 dex).

\subsection{Fitting Procedure}

Our fitting procedure is similar to that outlined in detail in
\citet{LBH05}. Briefly, we first normalize the flux from the
individual Balmer lines, in both the observed and predicted spectra,
to a continuum set to unity. This continuum is defined by fitting the
observed line profiles with a theoretical spectrum including a
polynomial with several free parameters to account for residual errors
in the flux calibration.  We have slightly improved the method by
using a larger wavelength range compared to previous analyses. In this
first step, the models are not used in any way to infer the
atmospheric parameters. This approach for normalizing the observed
spectra is used for $\Te > 16,000$ K and for $\Te < 9000$ K, while at
intermediate temperatures --- where the Balmer lines reach their
maximum strengths --- we use several pseudo-Gaussians profiles, which
constitute a more robust approach \citep{bergeron95}.  Finally, the
H$\alpha$ line is normalized independently from all other lines, using
theoretical line profiles in all cases.  We then proceed to a $\chi^2$
minimization between the observed and predicted line profiles,
convolved with a Gaussian instrumental profile with a resolution of 3
\AA\ (FWHM) appropriate for the SDSS spectra.

Our atmospheric parameters determinations for the DA stars in the SDSS
is meant to be completely independent of previous analyses, other than
the use of the DA designation from \citet{E06} for the main spectral
type in the definition of our sample. It is well known that there is
generally a cool and a hot solution for the fit to normalized Balmer
lines, centered on the maximum strength of the hydrogen lines near
$\Te \sim 13,500$ K. In a first iteration, we determine the
atmospheric parameters for each star assuming both a cool and a hot
solution, and adopting as a preliminary solution that which provides
the best agreement between the spectroscopic temperature $(T_{\rm
  spec})$ and the photometric temperature $(T_{\rm phot}$; the
photometric fits are described in Section 2.4).

The major improvement of our analysis is that we go through a second
iteration for all stars with ${\rm S/N}>12$ and perform a careful
visual inspection of each individual fit. While we only consider
normalized Balmer line profiles to determine the atmospheric
parameters, we also rely on two additional diagnostics to reclassify
and to flag interesting objects. The first diagnostic is the
dereddened $ugriz$ photometry, which we fit with the same model
fluxes. The second diagnostic is the direct comparison of the
synthetic and observed absolute fluxes, assuming that the SDSS spectra
are properly flux calibrated. Note that in the SDSS, the spectroscopic
and photometric observations are independent. These two additional
diagnostics are usually lacking in other large DA surveys, and this
certainly represents a significant advantage of the SDSS sample. Our
visual inspection revealed 52 confirmed magnetic white dwarfs, or
magnetic candidates, and we remove these objects from our sample
because these cannot be fitted, even approximately, with our
non-magnetic models. We refer to the studies of \citet{MAGN1},
\citet{MAGN2}, and \citet{MAGN3} for a more detailed analysis of these
stars. We also remove 3 objects that are most likely not white dwarfs
(J090917.09+002514.0, J204949.78+000547.3, and
J205455.83+005129.7). The second object has actually been reobserved
by \citet{kilic07} and confirmed as a non-degenerate. Finally we
remove two DA stars for which the SDSS observations cannot be fit with
our spectroscopic technique because more than two lines were
unavailable or unusable.

The most common flag in our analysis is the presence of a red flux
excess in 345 objects, generally with non-degenerate absorption lines,
which in almost all cases reveals the presence of an M dwarf
companion\footnote{We use the terminology of DA$-$M dwarf binaries
  throughout this work but it must be understood that some of them are
  merely candidates because the red excess is weak, and some
  companions are actually L dwarfs.}. For 65 of these systems, there
is an obvious contamination blueward of H$\beta$, and these are simply
too difficult to analyze with our spectroscopic approach. For the
remaining systems with a smaller contamination, we use a simple
technique that has proven to be sufficient in the present context. We
exclude one or two lines (H$\alpha$, H$\beta$) from the $\chi^2$ fit,
and in some cases the line cores when emission is present. We also
carefully inspected each fit, and were conservative in excluding all
possibly contaminated lines. We find out that in most cases, the
direct comparison of the synthetic and observed absolute fluxes is
good in the region where the lines are fitted. We also use the $u-g$
color index to help in discriminating between the cool or hot
solutions since the quality of the fit is generally a poor indicator
due to possible contamination.

Even if the SDSS contains mostly faint white dwarfs (see, e.g.,
Fig.~\ref{fg:f1}), there is still a large sample of high quality data
--- i.e. 665 spectra with ${\rm S/N}>30$, which is comparable in quality
and quantity to other large surveys of DA stars \citep[the PG survey for
instance;][]{LBH05}. In Figure \ref{fg:f2}, we show sample fits from these
high S/N observations. It is clear that by taking only the best
spectroscopic data in the SDSS sample, one can still construct one of
the best available sample of DA stars.

\subsection{$ugriz$ Photometry}

In our overall fitting procedure, we also make use of $ugriz$
photometric observations, which are available for most DA stars in the
SDSS-E06 catalog. We fit the observed photometry of each object by
using the grid of model atmospheres discussed in Section 2.2. We
removed all individual magnitudes with a flag indicating that there
might be a problem. Since the photometry is not sensitive to surface
gravity, we simply assume the spectroscopic $\logg$ value and fit only
the effective temperature $T_{\rm phot}$ and the solid angle $\pi
(R/D)^2$, where $R$ is the stellar radius and $D$ is the distance from
Earth. We apply a correction to the $u$, $i$, and $z$ bands of
$-0.040$, $+0.015$ and $+0.030$, respectively, to account for the
transformation from the SDSS to the AB magnitude system, as explained
in \citet{E06}. Finally, we deredden the photometry, in an iterative
fashion, using the distance of the star found from the previous
iteration, and the parameterization of \citet{harris06} for the amount
of reddening as a function of distance\footnote{We would like to point
  out here that interstellar absorption is assumed to be negligible
  for stars with distances $<100$~pc, and that the absorption is
  maximum for stars with distances from the Galactic plane $|z| >
  250$~pc. The absorption varies linearly along the line of sight
  between these distances.}.

In the course of our inspection of the photometric and spectroscopic
fits, we identified 37 double degenerates or helium-rich DA candidates
that will be discussed in Section 3.4 and 3.5\footnote{We also remove the
double lined binary J125733.63+542850.5 identified by
\citet{badenes09}, and the magnetic DAH+DB system J084716.21+484220.3
first identified by \citet{MAGN1}} together with a more general
comparison of spectroscopic and photometric temperatures. Note that we
do not make any direct use of the photometric temperatures since these
are intrinsically less accurate, with larger internal uncertainties
from the fitting procedure, and additional external uncertainties from
interstellar reddening and photometric calibration.

\section{RESULTS}

\subsection{Atmospheric Parameters}

We present in Table 1 the atmospheric parameters for our final sample
of 3072 DA stars with ${\rm S/N} > 12$ whose fits were visually
inspected. We exclude 161 objects, as discussed in the previous
section, and we postpone the analysis of 16 DAO stars to Section
4.3. When multiple spectra of the same star are available in the DR4,
we simply adopt the solution obtained with the highest signal-to-noise
spectrum. We convert $\logg$ values into stellar masses using the
evolutionary models with thick hydrogen layers of \citet{fontaine01}
below $\Te < 30,000$~K and of \citet{wood95} above this temperature;
we also provide in Table 1 the corresponding white dwarf cooling time
($\log\tau$). Low-mass white dwarfs, below 0.46 $M_{\odot}$ and $\Te <
50,000$~K, are likely helium core white dwarfs, and we rely instead on
evolutionary models from \citet{althaus01}. For masses higher than 1.3
$M_{\odot}$, we use the zero temperature calculations of
\citet{hamada61}. The absolute visual magnitude $M_V$ obtained from
spectroscopic measurements of $\Te$ and $\logg$ is also given for each
object following the calibration of \citet{holberg06}.

The errors of the atmospheric parameters given in Table 1 are the {\it
internal} uncertainties of our fitting procedure, which are
correlated to the S/N \citep[see, e.g.,][]{bergeron92}. The true error
budget also needs to take into account the {\it external}
uncertainties from the data calibration. There are 301 repeated
observations for stars in Table 1 that can be used to quantify the
external errors. These alternative observations, in general with S/N
only slightly lower than the primary observation, were fitted and
visually inspected with our standard procedure. In Table 2, we compare
the internal and external errors for these multiple observations. A
similar procedure has been used by \citet{E06}, but since
uncertainties are expected to be a function of both $\Te$ and S/N,
here we break the sample into various bins for each quantity. We find
that external uncertainties are always very similar to the internal
uncertainties. This confirms the consistency of the data acquisition
and 
pipeline reduction procedure of the SDSS spectra.
We note that our comparison cannot reveal systematic
reduction problems, however, and such effects will be discussed in Section 4.2
from a comparison with independent observations.

In Figure \ref{fg:f3}, we present the mass distribution as a function
of effective temperature for our sample of DA stars. As we want to
illustrate the effect of the quality of the observations on this
distribution, we show in the upper panel the results with the higher
quality spectra (${\rm S/N}>20$) and in the bottom panel with the
lower quality spectra ($20>{\rm S/N}>12$); each subsample has a
roughly equal number of stars. For clarity, we postpone the discussion
of the hottest stars ($\Te > 40,000$ K) to Section 4.3. Below
$\Te\sim13,000$~K, we observe the well-known increase in the mean
mass, the so-called high-$\log g$ problem discussed at length in
\citet{TB10} and references therein. The bottom panel indicates that
the mass distribution obtained with lower S/N data is similar to that
with higher quality spectra in the intermediate range of effective
temperature where the hydrogen lines are particularly strong (30,000 K
$> \Te > 10,000 $ K). However, larger spreads in the mass distribution
can been seen at the high and low temperature limits of our sample, in
a region where hydrogen lines become weaker and more sensitive to the
quality of the observations. Furthermore, since the flux in the blue
portion of the spectrum is intrinsically less important in cooler
stars, the S/N also becomes much smaller below 4000 \AA\ --- a region
sensitive to $\logg$ --- than at longer wavelengths, and the
atmospheric parameters of cooler stars are thus expected to become
less accurate. An examination of our error budget in Table 2 indeed
confirms that at cool temperatures and low S/N, the external error of
$\log g$ is indeed roughly twice that of the average of the sample.

As already mentioned, one of the most important aspects of our
analysis is the careful visual inspection of each individual fit. In
Figures \ref{fg:f4} and \ref{fg:f5} we show illustrative examples for
three objects.  We specifically look at the following diagnostics: (1)
the quality of the spectroscopic fit (left panels), (2) the quality of
the photometric fit (right panels), (3) the comparison of the
spectroscopic and photometric temperatures, (4) the superposition of
the observed photometry and fluxed spectra, and finally (5) the
superposition of the model fluxes normalized at $r$ and calculated at
the spectroscopic values of $\Te$ and $\logg$. On the top panels of
Figure \ref{fg:f4} is an example of a case with an excellent internal
consistency between model and spectroscopic/photometric observations,
a situation we observe for about 95\% of single DA stars. On the
bottom panels of Figure \ref{fg:f4} is an example of a problem we
encounter in about $5\%$ of the stars in our sample (although,
generally, to a lesser extent than what is shown here). In these
situations, the slope of the observed spectrum disagrees with that
inferred from photometric colors. This occurs more often for the
reddest and coolest white dwarfs in the sample. In most cases, all
other aspects of the visual inspection are as expected (i.e., a good
spectroscopic fit, and a good agreement between spectroscopic and
photometric temperatures). This clearly suggests that errors in the
flux calibration are the most likely explanation for this
discrepancy. The observed spectrum of J003511.63+001150.3 shown in
Figure \ref{fg:f4} is actually a repeated observation; a primary
spectrum with a slightly higher S/N was used to determine the
atmospheric parameters given in Table 1. The primary spectrum is
actually in agreement with the photometric colors, confirming the
hypothesis that a fraction of SDSS spectra may have an erroneous flux
calibration. \citet{kleinman04} claim that the SDSS spectra are
spectrophotometrically calibrated to within about 10$\%$, on average,
so some discrepancy, such as that shown in Figure \ref{fg:f4}, should
not be totally unexpected. The trend we observe, however, is that the
flux calibration is good to within a few percent for most objects, but
is significantly worse for a minority ($\sim$5\%) of cases. We added a
note in Table 1 (see Note 2) to flag these particular observations,
but these objects should not be considered as peculiar white
dwarfs. On the other hand, there are a few cases (about 1\%) where the
predicted energy distribution inferred from the spectroscopic solution
disagrees with both the slope of the observed spectrum {\it and} the
photometric colors, as illustrated on the top panel of Figure
\ref{fg:f5}. In other words, there is a poor match between $T_{\rm
  spec}$ and $T_{\rm phot}$, even though the spectroscopic fit appears
normal. We note that one could also select the cool spectroscopic
solution for this object, to obtain a better match with the
photometric temperature. This case is displayed on the bottom panel of
Figure \ref{fg:f5}, and while the agreement between $T_{\rm spec}$ and
$T_{\rm phot}$ is better, the Balmer line profiles are clearly at odds
with the model predictions. Objects similar to those shown here are
believed to be genuine peculiar stars and these are further discussed
in Sections 3.3 and 3.4.

\subsection{Mass distribution}

In this section, we attempt to best characterize the mass distribution
of DA stars using the SDSS-E06 sample. We restrict our analysis to
stars cooler than $\Te = 40,000$~K due to possible systematic effects
from metal contaminations (see Section 4.3), and to stars hotter than
13,000 K because of the high-$\log g$ problem discussed above. We also
eliminate, for the time being, all DA$-$M dwarf binaries; these will
be discussed further in Section 4.4. As a first step, we look for a
compromise between the quality of the spectra and the size of the
sample, as discussed in the Introduction. We saw in the preceding
section that we could obtain sound results using all stars with ${\rm
  S/N}>12$, but we should be cautious at the hot and cool ends of the
distribution.

To make a more quantitative assessment of the optimal sample for
computing the mass distribution, we break our sample into S/N bins
containing a nearly equal number of stars. In Figure \ref{fg:f6}, we
present the mean mass and standard deviation as a function of
S/N\footnote{For ${\rm S/N} < 12$, we used an automatic fitting
procedure similar to that employed for the higher quality spectra. In
order to get a relatively clean sample, we removed DA$-$M dwarf
composite systems using color criteria and all other outliers that did
not fall within the range of $6.5 < \log g < 9.5$}. In the ideal case
of a homogeneous spectroscopic data reduction as a function of S/N,
the mean mass should be constant over the whole sample. The standard
deviation, however, should be close to the intrinsic value of the
underlying mass distribution at high S/N, but should increase when the
errors of the individual mass measurements become significant compared
to the true dispersion of the sample. We can see in the upper panel of
Figure \ref{fg:f6} that the mean mass remains fairly constant, even
down to very low S/N values (${\rm S/N}\sim8$), which implies that as
a whole, we can trust the SDSS spectroscopic data in this range of
S/N. The standard deviation, displayed in the bottom panel of Figure
\ref{fg:f6}, appears as a more efficient means of selecting an
appropriate S/N cutoff. It shows clearly that for low S/N, the large
individual uncertainties in the mass measurements alter significantly
the mass dispersion. We conclude that a cutoff at ${\rm S/N} = 15$ is
a good compromise to obtain the best statistical significance, that
is, a large sample with sufficiently accurate spectroscopic
measurements. As can be seen from Figure \ref{fg:f1}, this cutoff
corresponds to a $g\sim19$ cutoff, similar to the value used by
\citet{kepler07} and \citet{degennaro08} based on analogous arguments.

In Figure \ref{fg:f7}, we show our final mass distribution for the DR4
sample (40,000 K $> \Te > 13,000 $ K), which yields a mean mass of
$\langle M\rangle = 0.613\ M_{\odot}$. We note the presence of an
excess of low-mass objects and a high-mass tail, which are qualitatively similar
to those reported by \citet{kepler07} in their analysis of the same
sample. We also demonstrate in this plot that the adoption of a
slightly larger S/N cutoff of 20, rather than 15, does not change
significantly the mean properties, or even the shape, of the mass
distribution. While it is obviously much more difficult to identify
subtypes (i.e., magnetics, DAZ, DAO, etc.) or to pinpoint the exact boundaries
of the ZZ
Ceti instability strip \citep{gianninas05}, the atmospheric parameters of normal
DA white dwarfs appear reasonable down to low S/N, with admittedly increasing
uncertainties.

\subsection{Comparison of Spectroscopic and Photometric Temperatures}

The comparison of our spectroscopic and photometric effective
temperatures is displayed in Figure \ref{fg:f8}. For clarity, we only
show the results for DA spectra with ${\rm S/N} > 20$; the comparison
with lower S/N data is similar, with a slightly higher
dispersion. Since the photometry becomes increasingly insensitive to
effective temperature at the hot end of the sample, we do not consider
objects with both $T_{\rm spec}$ and $T_{\rm phot}$ over 40,000 K. We
find that the agreement is generally good, with a small standard
deviation of $\sim$8\%, although some outliers are also clearly
present. More importantly, there is a significant offset, with the
spectroscopic temperatures being higher than the photometric
temperatures by $\sim$4\%, on average. There could be different
explanations for this discrepancy, which we discuss in turn.

One important uncertainty is the dereddening procedure, which has an
effect even for the brightest stars in the sample. The procedure used
here, taken from \citet{harris06}, seems realistic in terms of the
galactic structure, but the ratio of the actual to the full galactic
reddening along the line of sight is uncertain by at least a few
percent, due to the simplicity of the parameterization. The offset in
Figure \ref{fg:f8} increases to $\sim$6\% if we do not apply any
reddening correction to the photometry; conversely, a large and
unrealistic extinction, close to the maximum value, is necessary to
remove the offset at $\Te \sim 20,000$ K. The reddening uncertainty is
therefore not large enough to be the main source of the observed
offset.  The second possible explanation for the offset is related to
the photometric system transformations discussed in Section 2.4. Since
these corrections are based on older models and data reductions (from
Data Release 4), they might have to be revisited, although \citet{E06}
quote uncertainties of the order of $1\%$, smaller than the
discrepancy observed here. Furthermore, we verified that, on average,
the observed and predicted photometry are in agreement within less
than 0.01 magnitudes for all filters.

A third source of error might be related to a problem with the
spectroscopic temperatures. It is shown later in this work (Section
4.2) that the spectroscopic temperatures obtained using SDSS spectra
are on average 2\% higher (for $\Te > 13,000$ K) than those obtained
using data from \citet{gianninas11} for objects in common. This
discrepancy, likely due to systematic data reduction problems, could
easily explain half of the offset observed in Figure \ref{fg:f8}. A
fourth possible source of uncertainty is that we cannot claim to be
more accurate than 1\% in terms of the physics of the models, and
therefore on the absolute scale of the spectroscopic
temperatures. However, we can at least claim that our improved models
yield absolute visual magnitude measurements, derived from
spectroscopic values of $\Te$ and $\logg$, that are consistent with
those derived from trigonometric parallax measurements \citep[see
  Fig.~14 of][]{TB09}. Finally, another option would be that the
offset is real, at least for some objects. In the next section, we
demonstrate that unresolved DA+DA binaries may indeed cause a
systematic positive offset between spectroscopic and photometric
temperatures.

We conclude that each of the uncertainties discussed above are able to
explain about half of the observed offset, and there is thus no
easy way to interpret the absolute spectroscopic and photometric
temperature scales of white dwarfs at this point.  Until these issues
are resolved, the photometric temperature scale should not be
discarded too easily, even if the internal uncertainties are larger.

Finally, we find in the comparison of spectroscopic and photometric
temperatures displayed in Figure \ref{fg:f8} a small number of
outliers. A closer inspection of all objects with a 2$\sigma$
discrepancy reveals that some of these outliers (red circles in
Fig.~\ref{fg:f8}) have already been flagged in Table 1 as being
problematic due to poor spectrophotometry or glitches in the observed
spectra. These objects are thus unlikely to be real peculiar white
dwarfs. However, many other outliers exhibit He~\textsc{i} lines in
their spectra and/or have poor fits. This suggests that these objects
most likely represent DA+DB or DA+DC unresolved degenerate
binaries. We analyze these systems further in the following section.

\subsection{Double Degenerate Binaries}

To better understand, from an observational point of view, how double
degenerate binaries can be detected in the SDSS, we first present a
simulation performed using a set of synthetic models. Since we are
mostly interested here in the DA stars found in the SDSS, we simulate
double degenerate systems containing at least a DA star together with
another white dwarf of the DA, DB, or helium-rich DC spectral
type. Individual model fluxes are thus combined for each assumed
component of the system, properly averaged by their respective radius,
to which we add a Gaussian noise of ${\rm S/N} = 30$; synthetic
$ugriz$ photometry is also extracted from the resulting spectrum. The
simulated data are then analyzed with our standard fitting procedure
under the assumption of single DA star models. For the sake of
simplicity, all calculations are made for a sequence of $\Te$ from
6000 to 40,000 K with steps of 2000 K for both components of the
system. We computed all possible combinations in this range of
temperatures with equal values of $\log g = 8$ for both components, as
well as with a $\logg$ difference of 0.5 dex ($\log g$ = 7.75 and 8.25
to be explicit). The results of this experiment are presented in
Figure \ref{fg:f9} where we compare spectroscopic and photometric
effective temperatures; open circles represent DA+DA binaries, while
filled circles correspond to DA+DB/DC binaries.

An examination of our results first reveals that all DA+DA binaries
are indistinguishable from single DA stars in this diagram. Therefore,
one absolutely needs additional constraints, such as trigonometric
parallax or radial velocity measurements, to identify DA+DA binaries
in the SDSS. \citet{liebert91} have already shown that combining the
model spectra of two DA white dwarfs results, in general, in an
apparently normal object that can be fitted successfully with single
star models. What we find here is that the combined $ugriz$ photometry
can also be fitted with single star models, and that the photometric
temperature happens to be within $\sim$10\% of the spectroscopic
solution, even if the two DA components of the system have large
temperature differences. Figure \ref{fg:f9} also reveals that the
spectroscopic temperatures are systematically larger than the
photometric temperatures for DA+DA composites. A careful examination
of this puzzling result reveals that the offset is due to the fact
that the average spectroscopic temperatures just happen to be
systematically larger than the average photometric temperatures.
Since the observed offset in Figure \ref{fg:f8} is of exactly the same
order as that simulated in Figure \ref{fg:f9}, it is of course
tantalizing to suggest that most DA stars might actually be DA+DA
composite systems! But we will refrain from doing so since,
coincidentally, various systematic uncertainties in the spectroscopic
and photometric temperature determinations are expected to produce
similar offsets, as discussed in the previous section.

The situation is very different for the DA+DB/DC binary simulations
shown in Figure \ref{fg:f9}, for which a large number of outliers can
be easily identified. The spectroscopic fit of these objects would
also often be flagged due to the presence of He~\textsc{i} lines
and a poor fit of the lower Balmer lines (typically the depth of
H$\alpha$ is too shallow compared to a single star model). Even when the
temperature of the DB/DC component is only half that of the DA
component $(T_{\rm eff-DB/DC}/T_{\rm eff-DA} = 0.5)$, these objects
still appear as outliers in Figure \ref{fg:f9} since the $\sim$5\%
additional continuum flux is still important in the core of the deep
Balmer lines. For lower values of $T_{\rm eff-DB/DC}/T_{\rm eff-DA}$,
it becomes almost impossible to confirm the presence of a companion,
unless the He-rich component shows strong He~\textsc{i} lines and the
S/N is particularly high.

In light of these results, we refined our criteria to find all
DA+DB/DC in our sample by specifically searching for He~\textsc{i}
lines, but found only one additional star\footnote{We exclude here hot
  DAO stars with $\Te > 40,000$ K.} that was not in the $>2\sigma$
outlier region in Figure \ref{fg:f8}. In our simulation of double
degenerates, we also find a temperature regime ($T_{\rm spec} \sim
10,000$ K in Fig.~\ref{fg:f9}) where we obtain a good match between
photometric and spectroscopic temperatures, yet the spectroscopic fit
is poor at H$\alpha$ (and possibly additional low Balmer lines). One
example of such a system is shown at the bottom of Figure \ref{fg:f5},
where H$\alpha$ (and to a lesser extent H$\beta$ and H$\gamma$) is
clearly at odds with the model predictions, and appears {\it diluted}
by a companion. Six double degenerate candidates were identified in
this manner from our DA sample.

We now proceed to fit all our double degenerate candidates with
appropriate model spectra; our grid of DB models is described, for
instance, in \citet{limoges10} and references therein. We use an
alternative version of our fitting program described above to extract
the normalized hydrogen and helium (if present) line profiles of all
candidates. Then a $\chi^2$ minimization is performed between the
observations and our set of DA and DB/DC models, with the $T_{\rm
  eff}$ and $\log g$ of each component of the system considered a free
parameter. We assume that all objects are physical binaries, i.e.~at
the same distance, and therefore the flux ratio between the components
is fixed by the set of atmospheric parameters.

We find in our sample a total of 35 binary candidates, including 10
DA+DB and 25 DA+DC systems. The fits are displayed in Figures
\ref{fg:f10} and \ref{fg:f11}, and the corresponding atmospheric
parameters are reported in Table 3 along with additional comments for
each system. For all DA+DC candidates, it is not possible to fit all
four parameters since the dilution effects produced by a more massive
(smaller radius) DC component are qualitatively comparable to those of
a cooler object. Therefore, we simply assume a value of $\log g = 8$
for all DC stars. Furthermore, we observe that when the S/N is low, or
when the hottest object is a DB/DC star, it is difficult to constrain
the $\log g$ value of any component; in these cases a value of $\log g
= 8$ is assumed for both stars. We find that, in general, the errors
of the atmospheric parameters are significantly larger than those
obtained for single stars. This is a direct consequence of considering
two additional free parameters in our fitting procedure, and also
results from the weakness of some of the observed spectral features.

We also compared for all binary candidates the predicted and observed
$ugriz$ photometry. We show an example for DA+DB and DA+DC systems in
Figure \ref{fg:f12}. Here we calculate the predicted $ugriz$
photometry with the atmospheric parameters given in Table 3, and
simply match it to the observed photometry using a single scaling
factor. We thus make no attempt to actually fit the observed
photometry since there would be too many free parameters in our
fitting procedure. In general, we find that this straightforward
comparison between predicted and observed photometry yields an
acceptable match, which is better or similar to that obtained under
the assumption of a single star model, with the exception of a few
objects identified in Table 3. We note that, unlike single stars, the
predicted photometry is sensitive to the surface gravity of both stars
since the relative contribution of each component of the system to the
total flux --- and thus the shape of the energy distribution ---
depends on their respective stellar radius. Therefore we do not put
too much emphasis on small discrepancies, especially when the $\log g$
values are assumed in the spectroscopic fits.

All 35 binary systems listed in Table 3 represent new discoveries,
with the exception of J034229.97+002417.6 (also known as KUV
03399+0015) first identified as a DA+DB system by \citet{limoges10} as
part of their spectroscopic analysis of the white dwarfs found in the
Kiso survey. Our findings suggest that $\sim$1\% of all white dwarfs
are in compact DA+DB/DC double degenerate systems. We note, however,
that a binary system containing a fairly hot DB component is likely to
be classified as a DBA star in the DR4, and not be included in our
analysis. The mean $\log g$ value for the 20 DA components for which
the surface gravity could be constrained is $\langle \log g \rangle =
8.12$. Since most of these DA stars are in the range of effective
temperature where the high-$\log g$ problem is observed
($\Te\lesssim13,000$~K), this average value thus appears entirely
consistent with that of single DA stars. The mean surface gravity for
the 10 DB components is $\langle \log g \rangle = 8.21$, a value
significantly larger than that of DA stars. This is perhaps not
surprising since DB stars with $T_{\rm eff}\lesssim 15,000$ K (7 out
of 10 DB components in Table 3) have a tendency to have spectroscopic
$\log g$ values larger than average \citep[see,
e.g.,][]{beauchamp96,kepler07}.

We have also considered the alternative possibility that our composite
systems are single DAB stars. By computing a grid of DAB model spectra
with a homogeneous composition, and using the same physics as that
discussed above for the DA and DB models, we find that in all but one
case, the spectroscopic and photometric fits are less satisfactory
than for the DA+DB composite fits. This is not surprising since for a
DAB star, we expect a good match between $T_{\rm phot}$ and $T_{\rm
  spec}$, similar to a single DA star, but most of our DA+DB binary
candidates appear as outliers in Figure \ref{fg:f8}. The only object
for which we could not clearly distinguish between a DA+DB and a DAB
solution is J015221.12$-$003037.3, displayed in Figure
\ref{fg:f13}. 

For a few additional objects (see, e.g., J084742.22+000647.6,
J084958.32+093847.7 and J093432.66+065848.6), the core of the lower
Balmer lines (especially H$\alpha$ and H$\beta$) is poorly reproduced
by the models, despite a good overall spectroscopic fit and a decent
match to the observed photometry. One possible solution to this
problem is that the DB component might actually be a DBA star. The DBA
subtype is common enough to suggest that a few of them should be
hiding in our sample, and at least one DA+DBA composite system has
previously been identified in the MCT survey \citep{wesemael94}. Due
to higher atmospheric pressures in DBA stars compared to DA stars, the
hydrogen lines will be more quenched, or in other words, H$\alpha$ and
H$\beta$ are expected to be much stronger than the higher Balmer
lines. Hence, a DA+DBA composite system will first be identified from
unusually large equivalent widths of the lower Balmer lines. We
confirm, in Figure \ref{fg:f14}, that the use of a grid of DBA models
(with fixed H/He = $10^{-3}$ abundances) can slightly improve the
quality of the spectroscopic fits for the three objects mentioned
above. The addition of a free parameter --- the hydrogen abundance in
the DBA star --- could be used to improve the fits. However, it is not
easy to constrain, and we postpone any determination of this parameter
until independent, higher S/N observations are secured.

We find that cool, weakly magnetic white dwarfs, especially at low S/N,
often resemble DA+DC degenerate binaries since their lines appear
broader and weaker compared to non-magnetic DA stars. The photometric
fits of these magnetic stars are generally good, but a discrepancy
between the photometric and spectroscopic temperatures is also
present, obviously due to the fact that our non-magnetic models fail
to reproduce the observed spectrum. The presence of Zeeman-split line
cores can, in such cases, confirm the magnetic interpretation. However,
there is one published magnetic star candidate
\citep[J231951.72+010909.0;][]{MAGN2} which could easily fool our
fitting procedure, since the Zeeman splitting is not apparent, and the
{\it diluted} Balmer lines can be fitted almost perfectly with a DA+DC
composite, although with a poor photometric match. We have thus been
cautious to inspect every object that was initially flagged as a
possible DA+DC composite, and were able to eliminate six DA stars that
are more likely to be weakly magnetic stars rather than binaries
(J075842.68+365731.6, J132340.34+003338.7, J133007.57+104830.5,
J144244.18+002714.8, J150856.93+013557.0 and J224444.62+130521.5). All
these objects indeed show hints of triple or enlarged cores, and good
photometric fits can be achieved with single star models. We also
identify nine DA+DC candidates in Table 3, where the presence of a
weak magnetic field remains a valid option. In conclusion, it is now
clear that better S/N observations are needed to properly identify
DA+DC candidates.

Many binary candidates in our sample show very weak
hydrogen lines superposed on an otherwise flat spectrum, where the DC
star is obviously the hotter and more luminous component of the
system. These objects could alternatively be interpreted as single
helium-rich DA stars with only small traces of hydrogen in their
atmospheres. In principle, however, such stars could be differentiated
from degenerate binaries since the hydrogen lines in a
helium-dominated atmosphere would be heavily quenched by the high
photospheric pressure, with only H$\alpha$ and H$\beta$ visible in
their spectrum. In contrast, most hydrogen lines in the Balmer series
will be detectable in the spectrum of a lower pressure $\Te \sim 7000$
K DA atmosphere, diluted by a $\Te \sim 10,000$ K DC\footnote{We note
that the shape of H$\alpha$ is a poor diagnostic, since the dilution
by a DC produces an {\it apparent} broadening similar to the real
enhanced broadening of a dense helium-rich
atmosphere.}. Unfortunately, both scenarios predict hydrogen lines
that are extremely weak and difficult to detect unambiguously at low
S/N. We attempted to fit all our binary candidates with a grid of
mixed hydrogen and helium models and we find that in eight cases, the
fits are equally as good as the binary fits. The reason is that the
S/N is too low to detect the higher Balmer lines even if they were
present.

To conclude this section, we are fairly confident that we have
correctly identified nine DA+DB degenerate binaries in the SDSS-E06
sample, since the quality of the fits as well as the comparison
between the observed and predicted photometry clearly rules out the
possibility of a single DAB star. The picture is less obvious for our
DA+DC binary candidates (and one possible DAB star) but we are still
confident that most of the candidates listed in Table 3 are indeed
degenerate binaries. Further monitoring of these stars, such as radial
velocity measurements, will be required to confirm their binary
nature. Incidentally, all SDSS spectra are in fact taken in three or
more exposures, generally in the same night, to facilitate cosmic ray
rejection. \citet{badenes09} used these individual exposures to look
for large radial velocity shifts (120 km s$^{-1}$ or more) and found
one such object. There are 25 of our binary candidates that were also
examined as part of their SWARMS survey (C.~Badenes 2010, private
communication). In most cases, no obvious radial velocity shifts can
be detected. For J002322.44+150011.6, however, a strong cross
correlation is observed between individual exposures and there is a
hint of a large velocity shift although the lines are very weak. 
8-meter class telescopes will obviously be needed to achieve
good radial velocity measurements for our faint candidates.

\subsection{Helium-Rich DA White Dwarfs}

\citet{bergeron91} were the first to show on a quantitative basis
that the pressure effects produced on the hydrogen lines in a cool
($\Te\lesssim 12,000$~K) DA star with a high surface gravity could not
be distinguished from those produced by the presence of large amounts
of helium. The authors also suggested that the higher than average
$\log g$ values inferred from spectroscopic analyses of cool DA stars
could be the result of the presence of helium brought to the surface
by convective mixing. But the non-detection of He~\textsc{i} lines in
high-resolution Keck observations of cool DA stars by \citet{TB10}
ruled out the systematic presence of helium in these
atmospheres. However, there are rare white dwarfs, such as GD~362 and
HS 0146+1847, that were interpreted as massive DA stars, but the
detection of the weak He~\textsc{i} $\lambda5877$ line in
high-resolution spectroscopic data suggested instead that these stars
had normal masses with helium dominated atmospheres
\citep{koester05,zuckerman07}. Both of these white dwarfs are DAZ stars
with circumstellar disks, and the source of hydrogen and metals in
these helium-dominated atmospheres is likely due to accretion from a
water-rich asteroid \citep{jura09}.

Broad and shallow H$\alpha$ features are also observed in cool white
dwarfs, indicative of a trace of hydrogen (H/He $\sim10^{-3}$) in a
helium-dominated atmosphere \citep[see, e.g., Fig.~12
of][]{bergeron01}. The hottest of these objects is the DZA white dwarf
Ross 640 at $\Te\sim 8500$ K. While no circumstellar disk has been
reported for this object, the simultaneous presence of metals in its
spectrum again suggests an external source for
hydrogen. Alternatively, the presence of mixed H/He compositions could
be the result of the mixing of the thin superficial hydrogen
atmosphere with the deeper and more massive helium convection zone
\citep{TB08}. White dwarfs with mixed atmospheric compositions
become increasingly common at even cooler temperatures ($\Te\lesssim
6000$~K) where they can be easily identified from their strong
infrared flux deficiency due to collision-induced absorption by
molecular hydrogen \citep{kilic10}. 

We identified in our SDSS sample two cool and massive DA white dwarfs,
J090150.74+
091211.3 and J170204.81+593635.5, that we believe are
probably helium-rich objects. These stars were initially flagged as
$2\sigma$ outliers in the $T_{\rm phot}$ vs.~$T_{\rm spec}$ diagram
shown in Figure \ref{fg:f8}, but we were unable to fit them with
binary models. In both cases, the photometric temperature was
significantly larger than the spectroscopic value (by 23\% and 37\%,
respectively), and the $\log g$ value was unusually high (9.89 and
9.66, respectively). In Figure \ref{fg:f15}, we present our best fits
assuming helium-rich compositions. Because of the degeneracy between
surface gravity and helium abundance, we simply assume here a value of
$\log g=8$. For the photometric fits, both the surface gravity and the
hydrogen abundance are fixed, the latter set at the spectroscopic
value. We can see that the spectroscopic and photometric temperatures
are now in agreement to within $\sim$10\%, which shows that by relying
on $ugriz$ observations, one can partially break the degeneracy
between extremely high $\log g$ stars and normal mass helium-rich
objects.

The first object in Figure \ref{fg:f15}, J170204.81+593635.5, is also
a DAZ white dwarf with the obvious Ca~\textsc{ii} $\lambda$3969
blended with H$\epsilon$. It appears like a cooler counterpart of
GD~362 and HS 0146+1847, although with significantly smaller metal
abundances. It would be interesting to confirm if the simultaneous
presence of hydrogen and metals in the photosphere of this star could
be explained by the existence of a circumstellar disk. The second
object shown in Figure \ref{fg:f15}, J090150.74+091211.3, is even
cooler at $\Te \sim 8500$ K, and is similar to Ross 640 in terms of
temperature and hydrogen abundance, even though no metal lines are
detected in the spectrum of this object. These helium-rich white dwarf
candidates are also interesting because they bridge the temperature
gap between DAZ white dwarfs like GD~362 and Ross 640. Obviously,
higher S/N spectroscopic observations and trigonometric parallax
measurements would help to confirm our interpretation of these stars,
although these measurements for such faint ($g \sim 18.5$) objects are
admittedly difficult to secure.

\section{DISCUSSION}

\subsection{Reappraisal of Previous Analyses of the SDSS Data Release 4}

Prior to this work, individual atmospheric parameters for most stars
in the DR4 catalog could only be found in the original \citet{E06}
paper. These authors were careful to define the largest white dwarf
sample ever identified so far, but their spectroscopic analysis was
preliminary, and as emphasized by the authors, their \texttt{autofit}
program was only designed to offer a first-pass estimate of
temperatures and surface gravities and to flag outliers. Also, most
spectroscopic fits were not visually and individually inspected,
unlike in our analysis. The model atmospheres between both analyses
differ as well. In particular, our models rely on improved Stark
broadening profiles and we also account for NLTE effects at high
temperatures. Our atmospheric parameters should thus represent a
significant improvement over previous estimates, especially given the
latest improved DR7 data reduction \citep{kleinman09}.

In Figure \ref{fg:f16}, we compare our atmospheric parameters with
those obtained by \citet{E06}. The agreement is surprisingly good
considering the differences in the data reduction, model spectra, and
fitting procedures. We find very few outliers, which implies that both
fitting techniques are robust, the main exception being the DA$-$M
dwarf binaries, although Eisenstein et al.~stress that their
atmospheric parameters are appropriate only if the spectral
classification is DA or DB, without other subtle variations. We note
that our surface gravities are significantly lower at cool
temperatures ($\Te\lesssim 8500$~K, by as much as 1 dex), most likely
due to the fact that the models used by Eisenstein et al.~only include
Stark and Doppler broadening (see also Section 2 of \citealt{kepler07}
where similar models are used), while our models also take into
account neutral broadening. Finally, most subtypes identified here are
identical to those reported in Eisenstein et al. In addition to the
double degenerate candidates reported in Table 3, there are 36 objects
in Table 1 for which our classification differs from that of
Eisenstein et al. Most of them have faint or questionable features
(Zeeman splitting, M$-$Dwarf contamination and He \textsc{ii} or Ca
lines) and our different classifications might only result from
different thresholds in the detection of these features. We thus
conclude that the \texttt{autofit} program and method of analysis
employed by Eisenstein et al.~are reasonable, except that a more
careful visual inspection would identify additional subtypes and
problematic observations, which is an essential step in determining
individual parameters for these stars, and to compute accurate mean
values for the sample.

The second important study of the DA stars in the DR4 is that of
\citet{kepler07} who went further and mainly focused on the mass
distribution of the sample, neglecting subtypes (DA$-$M dwarf
binaries, magnetics, DAZ, and DAB stars), as we did in Section
3.2. They used a more extensive model grid than that of \citet{E06},
but the physics included in these models is similar. Even though
Kepler et al.~argue that the fits at low effective temperatures cannot
be trusted since this corresponds to the temperature range where the
high-$\log g$ problem is encountered, we believe that a proper
understanding of this problem begins with the best achievable analysis
of these cool DA stars. Furthermore, it is not explicitly stated in
\citet{kepler07} whether individual fits were visually examined in
order to define a clean sample of single white dwarfs.

The optimal way of comparing our mean mass determination to that of
\citet{kepler07} would be to start with a 1:1 study of individual
objects, to evaluate the differences in the models and sample
selections, as noted above, and also in the fitting
procedures\footnote{Kepler et al.'s fitting method includes the full
  spectrophotometric spectra, which they argue provides the lowest
  internal uncertainties. One should be cautious, however, with
  external uncertainties stemming from data reduction, such as those
  presented in Figure \ref{fg:f4} (bottom panel) and discussed in
  Section 3.1. While we do not claim that our fitting method is
  better, we suggest that it is equivalent, and more easily comparable
  to other DA surveys, which rely on an approach similar to
  ours.}. Unfortunately, since individual atmospheric parameters were
not published in their work, a detailed comparison is not possible
here, and only average values can be compared. We note that Kepler et
al.~still give the atmospheric parameters of some low-mass and
high-mass outliers identified in their mass distribution, a discussion
we postpone to Section 4.5.

We recomputed the mean mass of our sample in the range 100,000 K $>
\Te > 12,000$ K, i.e.~the same range of temperature used by
\citet{kepler07}, and we find the same value as before, $\langle
M\rangle = 0.61$ \msun, which can be compared to the value reported by
Kepler et al., $\langle M\rangle = 0.59$ \msun. This small difference
is entirely consistent with the shift produced by the use of our
improved Stark profiles, and it is comparable to the shift observed
for the PG sample \citep{TB09} where we found a significantly higher
mean mass (by $\sim$5\%) compared to calculations with older models.
This suggests that the effects of using different model spectra, data
reductions, sample selections, and fitting techniques, are small,
after all.

\subsection{Comparison with Bright DA White Dwarfs}

We now compare our results for the DA stars in the SDSS with the
ongoing spectroscopic survey of bright ($V\lesssim17$) DA white dwarfs
of \citet{gianninas11} drawn from the Villanova White Dwarf Catalog \citep{mccook99}.
The obvious advantage of this comparison is that we are using the same
model atmospheres and fitting technique in both analyses, and
therefore any disparity can be attributed to differences in
the sets of observations only. Unlike the SDSS spectra, however, the wavelength
coverage used in the Gianninas et al.~survey does not extend to
H$\alpha$. We thus repeated our analysis of the SDSS spectra without
including H$\alpha$ in our fitting procedure, and found nearly
identical results. We can therefore safely compare the results between
both surveys without further ado.

We find 89 stars in common between both surveys. These correspond of
course to the brightest DA stars in the SDSS, which were already known
prior to that survey. The comparison between both sets of atmospheric
parameters is displayed in Figure \ref{fg:f17}. We can see that the
SDSS temperatures are systematically larger by about 2\%, on average,
in the range 40,000 $ > \Te > $ 13,000 K, than those obtained using
the Gianninas et al.~spectra, while the $\log g$ values are lower by
about 0.06 dex, on average, in the same range of temperature, with
differences increasing at higher effective temperatures. As mentioned
above, since we are using the same theoretical framework and fitting
method, these differences can only be attributed to the use of
different spectra.

We compare in Table 4 the values of the mean mass, the standard
deviation, as well as the median mass for the SDSS sample, the
Gianninas et al.~sample, and the PG sample \citep{LBH05}; note that
the PG sample is a subset of the Gianninas et al.~sample. The
mean mass is quite sensitive to the number of low-mass and high-mass
outliers in each sample, while the median mass is more closely related
to the peak value of the mass distribution. For internal consistency,
we use the same range of effective temperatures to determine the
average properties of each sample. We also remove all magnetic stars,
double degenerates, and DA$-$M dwarf composite systems. We should mention
that the PG sample has already been analyzed in \citet{TB09} but the
values reported here differ slightly from those published as a result
of three minor improvements in our analysis: the addition of the
nonideal electronic perturbations in the equation of state (see
Section 2.2), our slightly different line normalization technique (see
Section 2.3), and our exclusion of DA subtypes, which all
contribute to lower the mean mass.

We first notice that the values of the mean mass and standard
deviation for the Gianninas et al.~sample are slightly larger than the
PG sample, although their median masses are nearly identical. This
suggests that there are a bit more high-mass outliers in the Gianninas
et al.~sample, which is not surprising since the Villanova White Dwarf
Catalog contains a large number of massive white dwarfs detected in
X-ray surveys \citep[see, e.g.,][]{vennes99}. Therefore, we feel it is
more appropriate, in what follows, to restrict our comparison to
median values only.

The results presented in Table 4 indicate that the median mass of the SDSS
sample is about 0.02 \msun\ lower than the PG and Gianninas et
al.~samples (corresponding to a $\log g$ value $\sim$0.04 dex
lower). This difference is of course consistent with the results shown
in the bottom panel of Figure \ref{fg:f17}, which compares $\log g$
values for objects in common between SDSS and Gianninas et al. We must
then conclude that the differences in the median values are mostly
likely due to problems in the data reduction rather than to selection
effects. Since the Gianninas et al.~sample includes white dwarf
spectra secured over many years using various telescopes and
instruments, systematic data reduction effects are less likely to be
present. Furthermore, problems with the calibration of the SDSS
spectra have been known to exist in the first few data releases
\citep{kleinman04,E06}. Even though these are believed to have been
fixed in the new Data Release 7 \citep{kleinman09}, we suggest here
that a small ($\sim$3\%) but systematic calibration problem may still
remain. As discussed in Section 3.3, a systematic shift in the
spectroscopic temperatures could also explain part of the problem observed
in the comparison with photometric temperatures.

Another way to look at this problem is to compare the mass
distribution as a function of effective temperature for both the SDSS
and Gianninas et al.~samples, as shown in Figure \ref{fg:f18}. We
first notice that the number of high-mass outliers is larger in the
Gianninas et al. sample due to the selection effects discussed above;
the most massive white dwarfs in this distribution are actually ROSAT
objects. We thus prefer to focus our attention on the peak of the
distributions; lines of constant mass at 0.55 $M_{\odot}$ and 0.70
$M_{\odot}$ have been included in Figure \ref{fg:f18} to guide the
eye. It is obvious that there is a systematic offset between both
distributions. The mass values in the SDSS increase from the hot end
of the sequence down to about 12,500 K, while they remain more
uniformly distributed around the mean in the Gianninas et
al.~sample. We note that the Gianninas et al.~distribution was shown
to be more uniform \citep{TB10} when using the improved model spectra
from \citet{TB09}. These results reinforce our conclusions that a
problem with the calibration of the SDSS spectra still exists, even in
the latest data release.

We finish this section with a short discussion of the cool end
($\Te\lesssim13,000$ K) of the mass distribution shown in the top
panel of Figure \ref{fg:f18}, where close to half of the SDSS stars
are located. We can clearly see an important increase in the mean mass
in this particular range of temperature, which corresponds to the well
known high-$\log g$ problem discussed at length by \citet{koester09}
and \citet{TB10}. These unrealistic mass values prevent us from going
much further in the analysis of these cool DA stars, but the mass
distribution can prove itself useful in better understanding the
nature of this problem. To better illustrate the shift in the mean
mass, we average in Figure \ref{fg:f19} the distribution into
temperature bins of 1000 K, and compute the corresponding mass
standard deviation. We observe that the SDSS mass distribution
exhibits an important and distinct triangular bump, with the mass
dispersion remaining fairly small and constant as a function of
temperature, even down to the very cool end of the distribution. The
fact that the mean mass appears to {\it decrease} again below 10,000 K
should be viewed with caution, however, since our model atmosphere
calculations include a free parameter in the treatment of the nonideal
effects from neutral particles, which allows some extra leverage to
``calibrate'' the $\log g$ values \citep[see Section 5.1
  of][]{TB10}. The SDSS sample definitely provides the clearest
picture so far of the high-$\log g$ problem, and it will certainly be
helpful in future investigations of this problem.

\subsection{Hot DA and DAO White Dwarfs}

We have postponed our discussion of the 184 hot DA stars ($\Te >
40,000$ K) until now due to the few extra steps needed to analyze
these objects. First of all, we uncovered 17 hot DAO stars in the
SDSS-E06 catalog, characterized by the He~\textsc{ii} $\lambda4686$
absorption line. All but two of these objects were previously
classified as DAO in the SDSS-E06 (the DAO J130815.21$-$015904.4 and
DAO+dM J094720.94+111734.7 were identified as DA and DA+dM,
respectively). We observe the so-called Balmer-line problem
\citep[see][and references therein]{gianninas10} in all of these DAO
stars, but also in many of our hot DA stars, especially at H$\alpha$
and H$\beta$. This problem manifests itself as an inability of the
model spectra to reproduce simultaneously all Balmer line profiles. A
solution to the Balmer-line problem was presented by \citet{werner96}
when they included carbon, nitrogen, and oxygen (CNO) -- with proper
Stark broadening -- in their model atmospheres. The main effect is a
cooling of the upper layers of the atmospheres, the formation region
of the core of the lower Balmer lines. These improved models are shown
to provide much better fits to the observed line profiles. More
reliable atmospheric parameters are then obtained, even if the CNO
metals are merely a proxy for all metals in the atmosphere and it is
by no means a determination of the CNO abundances. The 15 DAO stars
identified in the SDSS-E06 have already been analyzed with mixed H/He
models by \citet{hugelmeyer07}\footnote{Their sample also includes
  J163200.32$-$001928.3, clearly a DAO star but identified as a DO in
  the SDSS-E06 catalog, hence it was not selected in this work.} but
without metals included in the model calculations to account for the
Balmer-line problem.

We rely upon the NLTE model atmospheres with homogeneous H/He/CNO
compositions introduced by \citet{gianninas10} to fit the DAO stars in
our SDSS sample. These models, computed with TLUSTY and using improved
Stark broadening from \citet{TB09}, are similar to the pure hydrogen
models presented in Section 2.2 except for their chemical
compositions. The CNO metal abundances are fixed at the
\citet{asplund05} solar values. Our fitting procedure is very similar
to the one previously described in this work, except that we also fit
the helium abundance based on the profile of He~\textsc{ii}
$\lambda4686$. The improved atmospheric parameters for 16 DAO stars
are presented in Table 5. The DAO$-$dM composite system
J094720.94+111734.7 is instead fitted with regular DA models since the
H$\alpha$ and H$\beta$ lines are contaminated by the M dwarf, and the
parameters are given in Table 1. We also find that 27 hot DA stars in
SDSS show the Balmer-line problem (DA+BP). We fitted these objects
with the H/CNO models also discussed in \citet{gianninas10}; the
atmospheric parameters are reported in Table 1 (see Note 6). The
Balmer-line problem is resolved for the majority of the 43 DAO and
DA+BP objects fitted with CNO models.  Also, the fits without CNO tend
to underestimate surface gravities by $\sim$0.1-0.2 dex (see also
\citealt{gianninas10} for more details), hence the effects are
important on the properties of the hot end of the SDSS mass
distribution.

The size of our sample of hot DA white dwarfs is similar to that of
\citet{gianninas10}, who analyzed 152 DA stars above $\Te > 40,000$
K. The fraction of objects showing the Balmer line problem is roughly
the same, suggesting that this discrepancy can be efficiently detected
even at the lower average S/N of the SDSS sample. The additional
diagnostic that can be performed with H$\alpha$ in the SDSS helps in
identifying the Balmer-line problem at lower S/N. In the course of our
visual inspection, however, it was obvious that only extreme DA+BP
could be detected for some objects with S/N $<$ 20. The atmospheric
parameters of hot DA stars below this threshold should therefore be
viewed as a pure hydrogen approximation.

In Figure \ref{fg:f18}, both the SDSS and Gianninas et al. mass
distributions include the improved atmospheric parameters of objects
fitted with H/He/CNO models. We find a mean mass of $\langle M\rangle
= 0.54$ \msun~for our SDSS hot DA sample, which is significantly lower
than mean mass identified for cooler stars. This is expected, however,
considering that the data reduction problem identified in the previous
section is worse at the hot end of the distribution (see Figures
\ref{fg:f17} and \ref{fg:f18}). For the DAO stars, we obtain a mean mass of $\langle
M\rangle = 0.52$ \msun, a value very similar to the hot DAs, which
further confirms the suggestion of
\citet{gianninas10} that the DA and DAO share a common
history. If we make abstraction of the reduction problem, it is very
interesting to note that in both the SDSS and Gianninas et
al. surveys, the hot end of the mass distribution is a smooth
continuation of the sequence at cooler temperatures. This result is a
direct consequence of the new models computed in
\citet{gianninas10}, including NLTE effects, CNO metals (for a
majority of stars at $\log T_{\rm eff} > 4.8$) and improved Stark
broadened profiles.

\subsection{White Dwarf$-$M Dwarf Binaries}

We analyzed in our study a large number of DA$-$M dwarf binaries that
are part of the DR4 sample. Their atmospheric parameters are given in
Table 1, although these were not included in the calculations of the
mean properties of the SDSS sample in order to define a cleaner sample
of single DA stars. Most of these objects have already been analyzed
in detail by \citet{silvestri06} and \citet{heller09}, although both
studies were not necessarily restricted to the DR4 sample. One of the
reason is that the SDSS-E06 catalog was not meant to be complete in terms
of the white dwarf$-$M dwarf binary content. The colors of these
objects can be significantly different from those of single white
dwarfs, especially when the total flux is dominated by the M dwarf
(i.e., the colors are very red). In those cases, the object colors can
be close to the low-mass end of the main-sequence stellar locus, and
rejected by the cutoffs. In contrast, \citet{silvestri06} resorted to
an algorithm to automatically identify spectral features of white
dwarfs in all DR4 spectra in addition to the color selection. They
claim that their sample comprises almost all spectroscopically
observed DA$-$M dwarf in the SDSS up to Data Release 4.

The deconvolution technique developed by \citet{heller09} is the most
sophisticated in terms of the determination of the atmospheric
parameters of both the DA white dwarf and the main sequence
companion. They fit simultaneously the effective temperature and
surface gravity of both stars, as well as the metallicity of the M
dwarf component. However, the precision on the white dwarf parameters
is rather low, with uncertainties as large as 0.5 dex in $\log
g$. This approach is not really convenient in the present context
since we are mostly interested in the white dwarf component of the
system. \citet{silvestri06} developed a more approximate method to fit
the main sequence star by using a series of spectral templates with
average atmospheric parameters. Then, by iteration, they subtracted
the contribution of the M dwarf component until a clean white dwarf
spectrum is obtained. This procedure has some flaws since it is not
possible to completely remove all spectral features from the M dwarf
companion, but it still allows a larger number of Balmer lines to be
fitted in comparison with the contaminated spectrum.

We compare in Figure \ref{fg:f20} our atmospheric parameters ---
determined by simply excluding from the $\chi^2$ fit the Balmer lines
that are contaminated by the M dwarf companion --- with those of
\citet{silvestri06}. We find a very good agreement between both sets
of measurements, especially considering the differences in the model
atmospheres. Our values of $\Te$ and $\log g$ are slightly larger, on
average, than those of Silvestri et al., but this small offset is
entirely consistent with our use of the improved Stark profiles of
\citet{TB09}. These comparisons are similar to those shown in Figure
\ref{fg:f16} for single stars, which suggests that our method for
fitting DA$-$M dwarf binaries is fairly reasonable and comparable to
that of Silvestri et al. There are also a few outliers observed in
Figure \ref{fg:f20}, which are explained by the particular choice
between the cool or hot solutions for the DA star. It is a more
complicated choice when analyzing such systems, with any fitting
method, but we believe our identifications are good since our model
fluxes are well matched by the observed $u-g$ color index.

We finally show in Figure \ref{fg:f21} the mass distribution of the
DA$-$M dwarf binaries in our SDSS sample, which can be compared to
that for single stars displayed in Figure \ref{fg:f7}. The average
masses and the shape of both distributions are clearly different. It
is perhaps not surprising that the high-mass tail is absent from the
DA$-$M dwarf mass distribution since massive white dwarfs have
smaller radii and are less luminous, and they can thus be easily
overshadowed by their M dwarf companion. But even the peak of the mass
distribution appears at a slightly lower value than for single stars,
a conclusion that can also be reached by looking at Figure 6 of
\citet{silvestri06}. Given these selection effects, it is difficult to
conclude whether the observed differences are real, another reason why
we have refrained from including these white dwarfs in our computation
of the average properties of DA stars.

\subsection{Outstanding Objects}

Several extreme low-mass and high-mass white dwarfs have been
identified in the SDSS and these have been the subject of numerous
studies, including that of \citet[][see their Section 7]{kepler07}.
We find that our improved line profiles do not affect significantly
the results of these earlier analyses. The shift in $\log g$ of these
extreme mass white dwarfs is comparable to that of typical DA stars.
For example, we find a value of $\log g = 6.40$ for
J123410.36$-$022802.8, a value similar to that obtained by
\citet{liebert04}, $\log g = 6.38$, based on the same SDSS spectrum
and a similar fitting method. This object was later reobserved by
\citet{kilic07} at higher signal-to-noise ratio using the MMT, and we
simply refer to their analysis for a determination of improved
parameters for this star, as well as for other low-mass white dwarf
candidates identified in the SDSS (see their Table 1).

For extreme high-mass white dwarfs close to the Chandrasekhar limit,
the asymptotic relation between surface gravity and mass implies that
a typical change in the value of $\log g$ will result only in a
negligible change in mass. Therefore, we do not expect significant
changes compared to published values. And as it turns out, our mass
determinations for the most massive white dwarfs in our sample are in
the same range as those reported in Table 6 of \citet{kepler07}. We
should note, however, that some of them have been flagged in our Table
1 as problematic observations (e.g., J155238.21+003910.3 and
J110735.32+085924.5) and one should thus be cautious about their mass
determinations. Furthermore, we suspect that some of the lower S/N
candidates could be weakly magnetic stars. For instance,
J154305.67+343223.6, identified in Table 6 of Kepler et al., is
actually a magnetic white dwarf \citep{MAGN3}. We believe that in any case,
higher S/N observations are required to properly constrain the mass of
these stars (these objects are indeed extremely faint with $18<g<19$).

\section{CONCLUSION}

We presented an updated spectroscopic analysis of the DA white dwarfs
identified in the SDSS Data Release 4 catalog of \citet{E06}, with the
most recent data reduction from DR7, using our improved NLTE
model grid including Stark profiles with non-ideal gas effects
\citep{TB09,TB10}. A careful visual inspection of
each individual spectroscopic fit (S/N $>12$), together with a
comparison with $ugriz$ photometric fits, allowed us to obtain a
significantly cleaner sample and improved atmospheric parameters for
these DA stars compared to previous studies.

We also performed a simulation of DA+DA and DA+DB/DC double degenerate
binaries analyzed both photometrically and spectroscopically using
single DA star models. We showed that DA+DA unresolved binaries could
not be easily detected with the SDSS data, but that most DA+DB/DC
systems would appear as outliers when comparing spectroscopic
and photometric temperatures.  Using this approach, we were able to
identify 35 DA+DB/DC double degenerate candidates in the SDSS sample,
most of them discussed for the first time in our analysis. We find,
however, that it is rather difficult to confirm unambiguously our
interpretation of the binary nature of DA+DC candidates at low S/N
since these objects can easily be mistaken for magnetic white dwarfs or
helium-rich DA stars.

We evaluated that a lower cutoff at S/N = 15 in the computation of the
mean mass of DA white dwarfs provides the best statistical
significance for these stars in the SDSS sample.  Our calculations
yielded a mean mass of 0.613 $M_{\odot}$ compared to a value 0.593
$M_{\odot}$ previously reported by \citet{kepler07}. This difference
is entirely consistent with the shift of $\sim$0.03 $M_{\odot}$
expected from our improved models. We also compared our results for
bright DA stars in common between the SDSS survey and the Villanova White Dwarf
Catalog sample of \citet{gianninas11} using the same grid of model
atmospheres and fitting techniques. We unexpectedly found a mean mass
for this subsample that is significantly higher (by $\sim$0.03
$M_{\odot}$) in the Gianninas et al.~survey. We concluded that a small
problem with the data reduction still remains in the spectroscopic
calibration of the SDSS Data Release 7.

Since no white dwarf survey is as large in volume as the SDSS,
resolving this issue will be important to characterize the mass
distribution of DA stars using SDSS data. This will also help in
understanding the absolute temperature scale of DA white dwarfs, for
which an offset is actually observed between spectroscopic and
photometric temperatures. Inevitably, this work will also be
beneficial as a guide for the analysis of new objects identified in
the SDSS Data Release 7, which are likely to have properties very
similar to those identified here.

\acknowledgements
We thank Carles Badenes, Vincent Cardin, and Audrey Maiuro for their
contribution to this project. This work was supported by the
NSERC Canada and by the Fund FQRNT (Qu\'ebec).

\clearpage
 \clearpage
 \begin{deluxetable}{clcccccc}
 \tabletypesize{\scriptsize}
 \tablecolumns{8}
 \tablewidth{0pt}
 \tablecaption{SDSS DR4 Sample of DA White Dwarfs with S/N $>$ 12}
 \tablehead{
 \colhead{SDSS name} &
 \colhead{Plate$-$MJD$-$Fiber} &
 \colhead{$\Te$ (K)} &
 \colhead{$\logg$} &
 \colhead{$M/M_{\odot}$} &
 \colhead{M$_{V}$} &
 \colhead{$\log \tau$} &
 \colhead{Notes}
 }
 \startdata
     J000006.75$-$004653.8&0685-52203-225&  10850 (160)& 8.39 (0.10)& 0.85 (0.07)& 12.52&  8.95&     \\
     J000022.53$-$105142.1&0650-52143-217&   8620 (110)& 8.31 (0.15)& 0.79 (0.10)& 13.24&  9.18&1    \\
     J000022.87$-$000635.7&0387-51791-166&  23010 (470)& 7.44 (0.06)& 0.42 (0.01)&  9.57&  7.59&     \\
     J000034.07$-$010819.9&0685-52203-187&  13090 (220)& 8.01 (0.05)& 0.61 (0.03)& 11.52&  8.50&     \\
     J000104.05+000355.8&0685-52203-490&  13710 (600)& 8.06 (0.08)& 0.64 (0.05)& 11.51&  8.50&     \\
     J000127.48+003759.1&0685-52203-491&  18560 (490)& 7.84 (0.09)& 0.53 (0.04)& 10.66&  7.88&     \\
     J000308.32$-$094147.0&0650-52143-550&   8690 (90)& 8.50 (0.12)& 0.92 (0.08)& 13.53&  9.36&     \\
     J000357.63$-$004939.1&0387-51791-005&   9740 (100)& 8.95 (0.10)& 1.18 (0.04)& 13.98&  9.41&     \\
     J000428.98+005801.9&0685-52203-621&  16410 (500)& 7.85 (0.10)& 0.54 (0.05)& 10.90&  8.10&     \\
     J000441.75+152841.1&0751-52251-393&   8710 (60)& 8.20 (0.09)& 0.72 (0.06)& 13.02&  9.07&     \\
     J000622.61+010958.7&0388-51793-448&  39040 (1930)& 7.73 (0.23)& 0.54 (0.09)&  9.03&  6.69&     \\
     J000630.56+002323.9&0388-51793-424&  23590 (770)& 7.89 (0.11)& 0.58 (0.06)& 10.30&  7.45&     \\
     J000636.61+160237.7&0751-52251-528&   9620 (100)& 8.30 (0.12)& 0.79 (0.08)& 12.80&  9.03&     \\
     J000716.84$-$101908.4&0651-52141-230&  20150 (740)& 7.87 (0.11)& 0.55 (0.06)& 10.56&  7.75&     \\
     J000737.18$-$090629.3&0651-52141-416&  20240 (410)& 7.88 (0.06)& 0.56 (0.03)& 10.56&  7.71&     \\
     J000738.03+004003.3&1490-52994-507&  10460 (70)& 8.32 (0.06)& 0.81 (0.04)& 12.53&  8.95&     \\
     J000946.45+144310.6&0751-52251-101&  24570 (240)& 7.89 (0.03)& 0.57 (0.02)& 10.22&  7.31&     \\
     J001038.78$-$003241.5&0388-51793-074&  10240 (120)& 8.47 (0.11)& 0.90 (0.07)& 12.86&  9.10&     \\
     J001148.19$-$092110.2&0652-52138-348&  12680 (180)& 7.84 (0.06)& 0.52 (0.03)& 11.34&  8.43&     \\
     J001245.60+143956.4&0752-52251-221&  11270 (120)& 7.89 (0.07)& 0.54 (0.04)& 11.65&  8.60&     \\
     J001339.19+001924.3&0389-51795-431&   9600 (10)& 8.32 (0.02)& 0.80 (0.01)& 12.84&  9.04&2    \\
     J001415.59$-$103505.8&0651-52141-023&   9920 (90)& 8.42 (0.09)& 0.86 (0.06)& 12.87&  9.09&     \\
     J001427.04+135058.4&0752-52251-135&   8870 (60)& 8.32 (0.08)& 0.80 (0.05)& 13.15&  9.14&     \\
     J001448.82+002027.3&0687-52518-468&   8960 (90)& 8.08 (0.13)& 0.64 (0.08)& 12.72&  8.97&2    \\
     J001518.88+135332.8&0752-52251-099&   8590 (30)& 8.20 (0.04)& 0.72 (0.03)& 13.08&  9.09&     \\
     J001556.07$-$000515.3&0389-51795-493&   9300 (80)& 8.27 (0.10)& 0.77 (0.06)& 12.87&  9.04&     \\
     J001629.05$-$004451.0&0687-52518-131&   9640 (70)& 8.30 (0.09)& 0.79 (0.06)& 12.78&  9.02&     \\
     J001643.36+152410.8&0752-52251-584&   8060 (60)& 8.29 (0.09)& 0.78 (0.06)& 13.46&  9.25&     \\
     J001654.67+151432.1&0752-52251-590&  12870 (420)& 8.23 (0.09)& 0.75 (0.06)& 11.87&  8.68&     \\
     J001655.51$-$005604.5&0687-52518-100&  16020 (270)& 7.94 (0.05)& 0.58 (0.03)& 11.07&  8.18&     \\
     J001747.15+155204.4&0752-52251-606&  11800 (300)& 8.35 (0.10)& 0.83 (0.07)& 12.23&  8.85&     \\
     J001749.24$-$000955.5&0687-52518-109&  54900 (1640)& 7.47 (0.11)& 0.50 (0.03)&  8.11&  6.11&3    \\
     J001836.15+003151.4&0688-52203-348&  11850 (120)& 8.12 (0.04)& 0.68 (0.03)& 11.85&  8.67&2    \\

 \enddata
 \tablecomments{Table 1 is available in its entirety in the electronic
 edition of the {\it Astrophysical Journal}. A portion is shown here
 for guidance regarding its form and content. (1) Incomplete
 wavelength coverage or glitches in the Balmer lines; (2)
 poor match between the slope of the observed spectra and the $ugriz$
 photometry; (3) main-sequence companion, one or two lines removed
 from the fit, and the line cores also removed due to emission; (4)
 same as previous but with no emission; (5) flagged in our visual
 inspection with a poor fit but no clear explanation was found; the
 atmospheric parameters under the assumption of a normal single DA
 should be regarded with caution; (6) DA with the Balmer-line problem
 and models with CNO were used; (7) emission in the line cores but no
 red excess; (8) DAZ; (9) DAO+dM.  For one object where the fitted
 $\Te$ exceeds our models grid limit, we fix the parameters at $\Te =
 140,000$ K and $\log g = 8$.}
 \end{deluxetable}

\clearpage
\clearpage
\begin{deluxetable}{lccccc}
\tablecolumns{6}
\tablewidth{0pt}
\tablecaption{DA White Dwarfs with Multiple Measurements}
\tablehead{
\colhead{S/N range} &
\colhead{N$^{a}$} &
\colhead{$\langle \sigma_{\rm Teff} \rangle^{b}$ (\%)} &
\colhead{$\langle \sigma_{\rm Teff-multiple} \rangle^{c}$ (\%)} &
\colhead{$\langle \sigma_{\rm \log g}\rangle^{b}$} &
\colhead{$\langle \sigma_{\rm \log g-multiple} \rangle^{c}$} 
}
\startdata
\cutinhead{$T_{\rm eff} < 13,000$ K}
$12 < \langle{\rm S/N}\rangle < 15$ &  18 & 1.3 & 1.0 & 0.12  & 0.14 \\
$15 < \langle{\rm S/N}\rangle < 20$ &  31 & 1.2 & 1.0 & 0.10  & 0.14 \\
$\langle{\rm S/N}\rangle > 20$      &  57 & 0.7 & 0.7 & 0.05  & 0.07 \\
\cutinhead{$T_{\rm eff} > 13,000$ K}
$12 < \langle{\rm S/N}\rangle < 15$ &  10 & 3.2 & 1.3 & 0.11  & 0.06 \\
$15 < \langle{\rm S/N}\rangle < 20$ &  24 & 2.7 & 2.6 & 0.08  & 0.07 \\
$\langle{\rm S/N}\rangle > 20$      &  54 & 1.8 & 1.8 & 0.05  & 0.05 \\
\enddata
\tablenotetext{a}{Number of stars with multiple measurements. Subtypes (DAO, DAZ, DA+dM and
  DA+DB) are not considered.}  \tablenotetext{b}{Mean internal
  uncertainty from the fitting procedure.}  \tablenotetext{c}{Average
  of the standard deviation between multiple observations.}
\end{deluxetable}

\clearpage
\clearpage
\begin{deluxetable}{clccccc}
\tabletypesize{\scriptsize}
\tablecolumns{7}
\tablewidth{0pt}
\tablecaption{DA+DB/DC Binary Candidates}
\tablehead{
\colhead{SDSS name} &
\colhead{Plate-MJD-Fiber} &
\colhead{$\Te$ (K) H-comp} &
\colhead{$\logg$ H-comp} &
\colhead{$\Te$ (K) He-comp} &
\colhead{$\logg$ He-comp} &
\colhead{Notes}
}
\startdata
J002322.44+150011.6&0753-52233-177&  6490 (420)&8 & 10750 (390)&8 &1,2       \\
J015221.12$-$003037.3&0402-51793-114& 34170 (1640)& 7.87 (0.14)& 14400 (3580)& 8.05 (0.95)&3         \\
J034229.97+002417.6&0714-52201-551& 13420 (170)& 8.32 (0.04)& 14380 (160)& 8.40 (0.06)&4         \\
J040852.23$-$045504.7&0465-51910-482&  7610 (200)&8 & 10730 (380)&8 &2,5       \\
J074436.39+451534.4&1737-53055-281&  9640 (230)&8 &  6770 (580)&8 &6         \\
J080743.57+442148.0&0439-51877-188&  8140 (170)& 8.39 (0.14)&  5380 (420)&8 &5,6       \\
J082403.91+310448.0&0931-52619-558&  6080 (220)&8 & 10340 (210)&8 &1,2       \\
J083353.71+385218.8&0828-52317-277& 26100 (1250)& 8.01 (0.10)& 16640 (910)& 7.86 (0.17)&5         \\
J084226.97+374040.0&0864-52320-524& 12330 (300)& 7.98 (0.20)&  9030 (900)&8 &          \\
J084608.19+053818.0&1187-52708-630&  9500 (260)& 8.21 (0.16)& 10040 (820)&8 &6         \\
J084742.22+000647.6&0467-51901-052& 11840 (320)& 8.19 (0.11)& 17940 (790)& 8.79 (0.15)&7         \\
J084958.32+093847.7&1760-53086-411& 10110 (320)& 8.35 (0.13)& 10410 (570)&8 &5,6,7     \\
J085159.31+532540.3&0449-51900-311&  7120 (270)&8 & 10830 (370)&8 &1,2       \\
J085858.49+301231.3&1590-52974-148& 12570 (290)& 8.16 (0.09)& 13830 (320)& 8.33 (0.13)&          \\
J093432.66+065848.6&1196-52733-093& 38700 (1620)& 7.52 (0.08)& 27870 (9140)& 7.71 (0.28)&5,7       \\
J093944.64+371617.8&1275-52996-630& 12800 (890)& 8.05 (0.13)& 13670 (530)&8 &          \\
J095902.54+451110.4&0942-52703-455&  7060 (730)&8 &  8770 (2730)&8 &2         \\
J102003.38+000902.6&0271-51883-557&  7120 (240)&8 & 10080 (710)&8 &1,2       \\
J102626.01+135745.0&1747-53075-430&  7800 (230)&8 & 10650 (510)&8 &1,2       \\
J122143.98+590747.8&0955-52409-041& 10390 (260)& 7.99 (0.36)&  9210 (1730)&8 &6         \\
J131907.32$-$023406.4&0341-51690-266&  9520 (300)& 8.29 (0.23)& 14330 (630)& 8.61 (0.30)&          \\
J134259.26+530519.2&1042-52725-076&  6420 (340)&8 & 11280 (220)&8 &1,2       \\
J140600.55+643312.9&0498-51984-163& 11280 (310)& 8.07 (0.17)& 12650 (470)& 8.11 (0.28)&          \\
J141005.74$-$023500.2&0916-52378-266&  7610 (470)&8 &  9870 (1760)&8 &1,2       \\
J141516.10$-$010912.1&0303-51615-057&  8190 (170)&8 & 10650 (440)&8 &6         \\
J152145.91+393128.0&1293-52765-385&  7070 (560)&8 &  9750 (2380)&8 &1,2       \\
J154710.83+442848.1&1333-52782-119& 10700 (160)& 8.24 (0.12)&  6520 (510)&8 &6         \\
J162757.07+331346.1&1058-52520-221& 10020 (150)& 8.13 (0.16)&  6710 (650)&8 &6         \\
J164306.05+442638.0&0629-52051-493&  7740 (180)&8 & 10640 (390)&8 &          \\
J172037.26+271914.8&0979-52427-501& 11840 (500)& 8.23 (0.17)& 14360 (530)& 8.31 (0.25)&          \\
J204036.43$-$001004.1&0981-52435-139&  7760 (180)&8 & 11350 (280)&8 &2         \\
J210155.81$-$005745.0&0984-52442-049& 12040 (180)& 7.59 (0.07)& 12340 (360)& 7.96 (0.15)&5         \\
J213819.85+112311.3&0731-52460-632&  8810 (190)& 8.57 (0.17)&  7680 (770)&8 &5         \\
J223437.86+002111.7&0673-52162-630& 12390 (560)& 8.25 (0.19)&  8370 (1000)&8 &5,6       \\
J224430.36+133430.2&0740-52263-264& 11210 (300)&8 & 15160 (400)&8 &          \\
\enddata
\tablecomments{(1) Helium-rich DA an equally valid possibility; (2) very faint H lines, difficult to fit in any cases; (3) could also be a DAB; (4) also in \citet{limoges10}; (5) average photometric match; (6) weakly magnetic also a possibility, good photometric fit to a single star; (7) cores of the H$\alpha$ and H$\beta$ lines poorly fitted.}
\end{deluxetable}

\clearpage
\clearpage
\begin{deluxetable}{cccccc}
\tablecolumns{6}
\tablewidth{0pt}
\tablecaption{Mean Properties of DA White Dwarf Samples}
\tablehead{
\colhead{Sample} &
\colhead{$\langle M/M_{\odot} \rangle$} &
\colhead{Dispersion} &
\colhead{Median Mass} &
\colhead{Sample Description}
}
\startdata
SDSS & 0.613 & 0.126 & 0.594 & \citet{E06} \\
Palomar-Green & 0.629 & 0.128 & 0.610 & \citet{LBH05} \\
White Dwarf Catalog & 0.638 & 0.143 & 0.610 & \citet{gianninas09} \\
\enddata
\end{deluxetable}

\clearpage
 \clearpage
 \begin{deluxetable}{ccccccccc}
 \tabletypesize{\scriptsize}
 \tablecolumns{8}
 \tablewidth{0pt}
 \tablecaption{SDSS DR4 Sample of DAO White Dwarfs with S/N $>$ 12}
 \tablehead{
 \colhead{SDSS name} &
 \colhead{Plate$-$MJD$-$Fiber} &
 \colhead{$\Te$ (K)} &
 \colhead{$\logg$} &
 \colhead{$\log$ He/H} &
 \colhead{$M/M_{\odot}$} &
 \colhead{M$_{V}$} &
 \colhead{$\log \tau$} &
 }
 \startdata
     J034831.33+004616.3&1242-52901-412&  90730 (2980)& 7.15 (0.11)&$-$2.12 (0.18)& 0.54 (0.02)&  6.76&  4.78 \\
     J081618.79+034234.1&1184-52641-171&  51740 (3550)& 7.53 (0.30)&$-$1.06 (0.39)& 0.51 (0.10)&  8.30&  6.31 \\
     J082705.53+313008.2&0932-52620-126&  78550 (2280)& 7.33 (0.08)&$-$2.41 (0.13)& 0.54 (0.02)&  7.35&  5.38 \\
     J101015.59+115711.3&1745-53061-218&  52500 (2820)& 7.38 (0.27)&$-$0.68 (0.22)& 0.47 (0.08)&  8.00&  6.02 \\
     J120927.93$-$030206.2&0332-52367-184&  79530 (4870)& 7.13 (0.16)&$-$1.95 (0.25)& 0.50 (0.04)&  6.91&  4.93 \\
     J121743.11+623118.2&0779-52342-152&  98170 (5220)& 6.98 (0.20)&$-$1.19 (0.19)& 0.53 (0.04)&  6.26&  4.35 \\\
     J125029.51+505317.3&1279-52736-450&  71660 (13240)& 7.08 (0.40)&$-$1.03 (0.61)& 0.46 (0.11)&  6.98&  5.04 \\
     J130815.21$-$015904.4&0340-51691-358&  53040 (1480)& 7.73 (0.10)&$-$2.49 (0.25)& 0.58 (0.04)&  8.64&  6.31 \\
     J131925.92+531715.0&1040-52722-015&  96320 (16210)& 6.73 (0.53)&$-$1.19 (0.52)& 0.48 (0.11)&  5.73&  3.33 \\
     J135356.88$-$025630.4&0914-52721-214&  53290 (1890)& 7.76 (0.13)&$-$2.15 (0.28)& 0.59 (0.05)&  8.68&  6.30 \\
     J145606.73+491116.5&1048-52736-619&  93050 (6230)& 6.64 (0.18)&$-$1.15 (0.19)& 0.45 (0.05)&  5.61&  3.78 \\
     J153102.39+534900.6\tablenotemark{a}&0616-52442-320&  78680 (7970)& 6.85 (0.25)&$-$1.96 (0.40)& 0.44 (0.06)&  6.35&  4.16&  \\
     J160236.07+381950.5&1055-52761-473&  84630 (5290)& 7.03 (0.17)&$-$2.14 (0.28)& 0.49 (0.04)&  6.62&  4.57  \\
     J161441.98+370548.1&1056-52764-546&  55640 (1710)& 7.66 (0.10)&$-$3.22 (0.34)& 0.56 (0.04)&  8.45&  6.24   \\
     J170508.81+212019.2&1425-52913-570&  50300 (1000)& 7.67 (0.08)&$-$2.70 (0.24)& 0.55 (0.03)&  8.59&  6.36   \\
     J235137.23+010844.2&0684-52523-370&  89500 (8920)& 7.56 (0.33)&$-$1.39 (0.40)& 0.63 (0.09)&  7.65&  5.59   \\
 \enddata
 \tablenotetext{a}{Observational glitch.}
 \end{deluxetable}

\clearpage

\figcaption[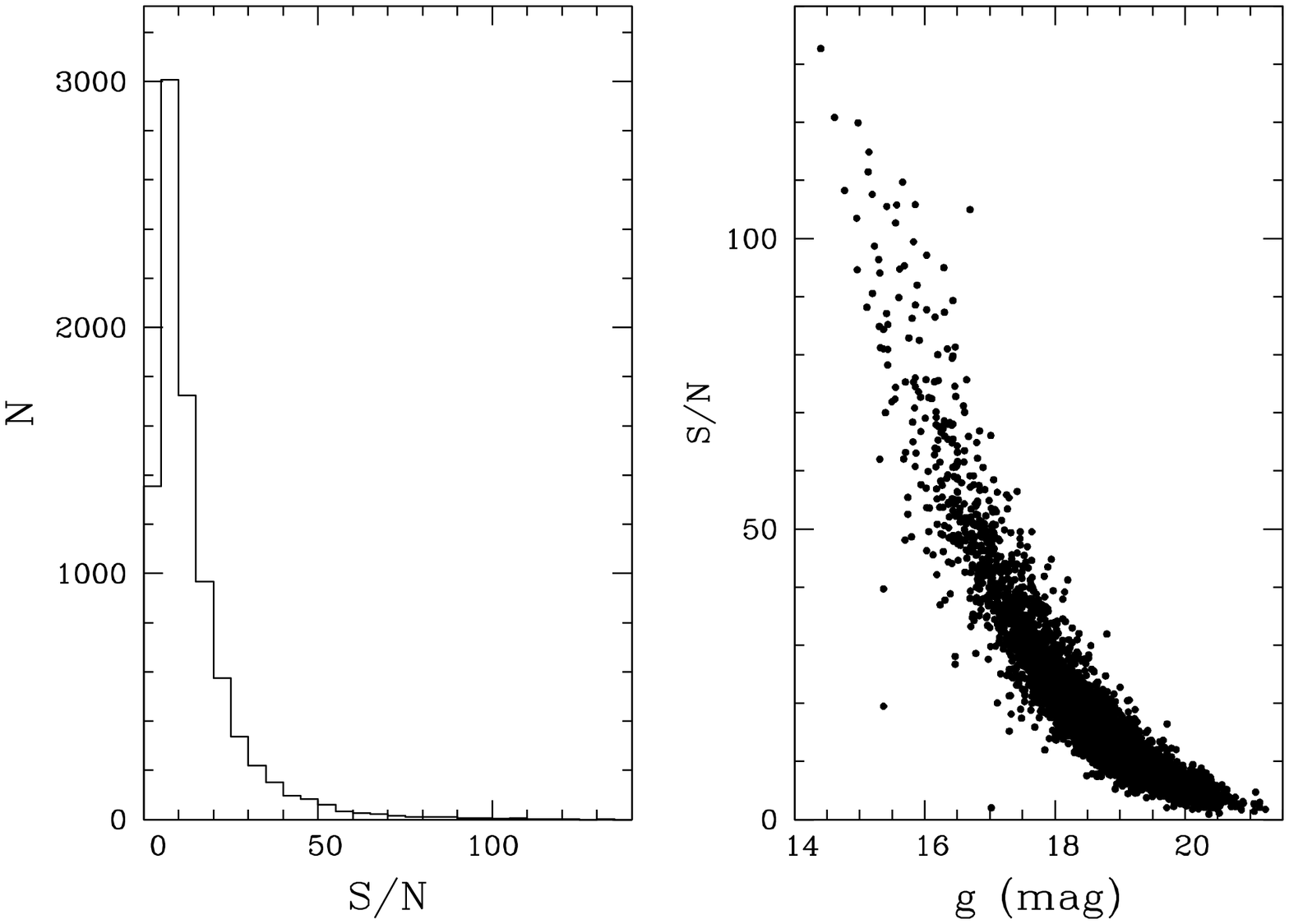] {Left panel: distribution of S/N for
all 8717 DA spectra in the SDSS-E06 sample. Right
panel: S/N as a function of the observed $g$ magnitude.\label{fg:f1}}

\figcaption[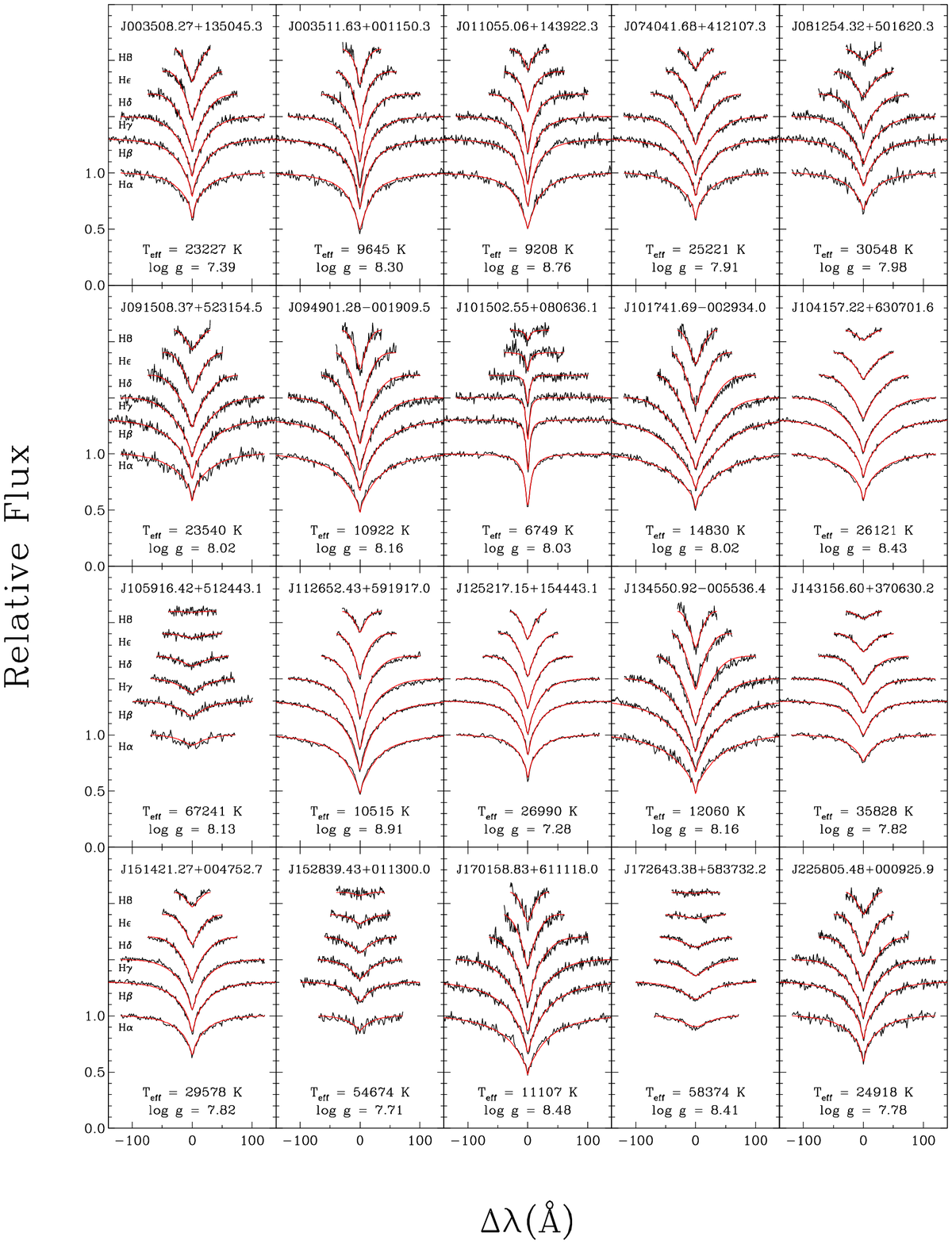] {Sample fits for 20 DA stars in the SDSS sample
with high signal-to-noise spectroscopic data (${\rm S/N}>30$). The
atmospheric parameters are given in each panel.
\label{fg:f2}}

\figcaption[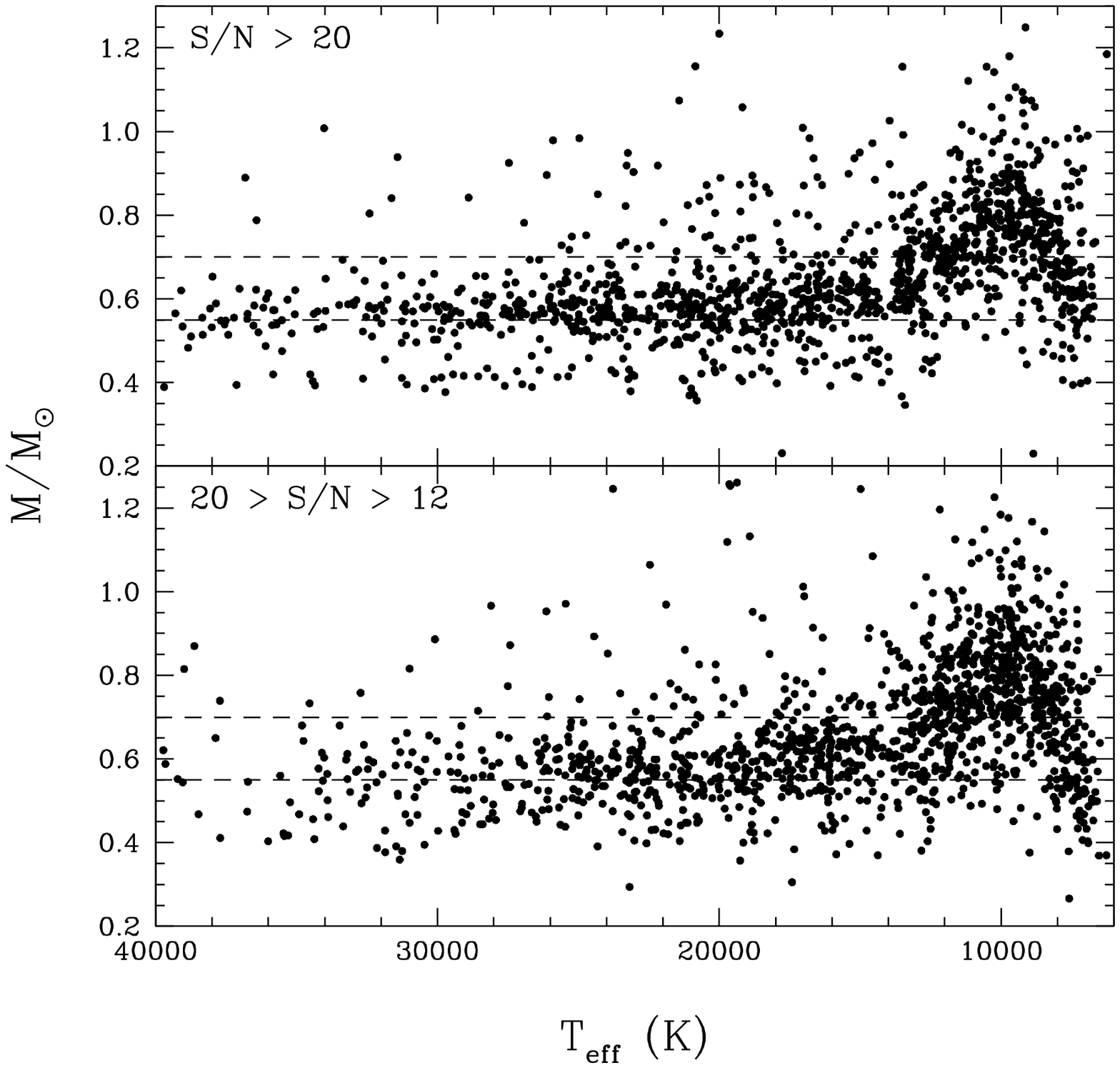] {Mass distribution as a function of effective temperature
for DA stars in the SDSS with $\Te < 40,000$~K, for 
two separate ranges of S/N. Lines of constant mass at 0.55 $M_{\odot}$ and
0.70 $M_{\odot}$ are shown as a reference.
\label{fg:f3}}

\figcaption[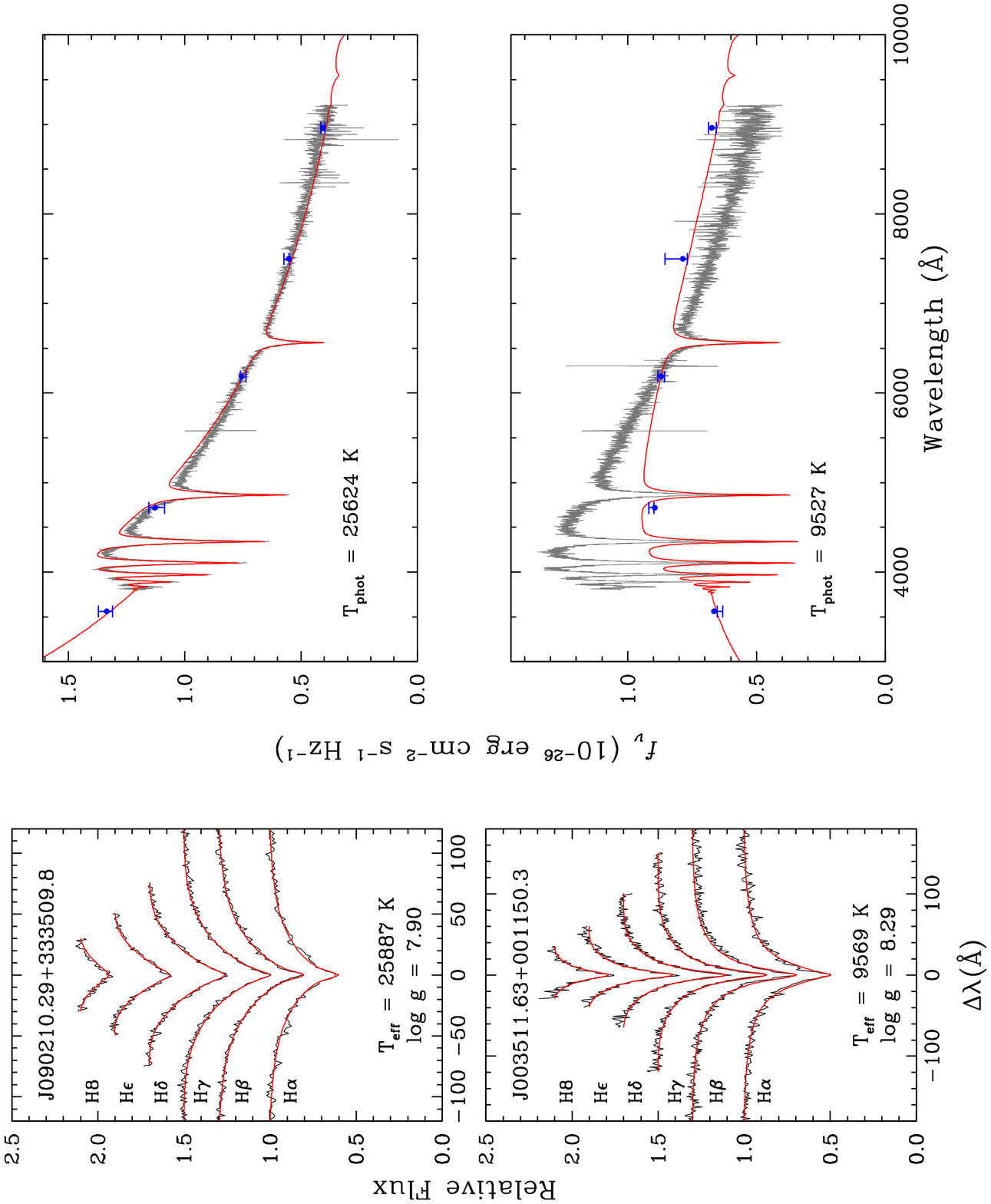] {Sample fits of DA stars in the SDSS sample with
  the atmospheric parameters given in each panel (left panels). In the
  right panels, we show complementary data used in our visual
  inspection. The y-axis scale is fixed by the observed photometric
  fluxes (blue error bars) that are fitted with average model fluxes
  (blue filled dots). The photometric temperature is given in each
  panel. Both the fluxed spectra (in gray) and synthetic model fluxes
  (in red), calculated at the spectroscopic parameters given in the
  left panel, are then scaled to the $r$ photometric band. The
  objects are discussed in the text.
\label{fg:f4}}

\figcaption[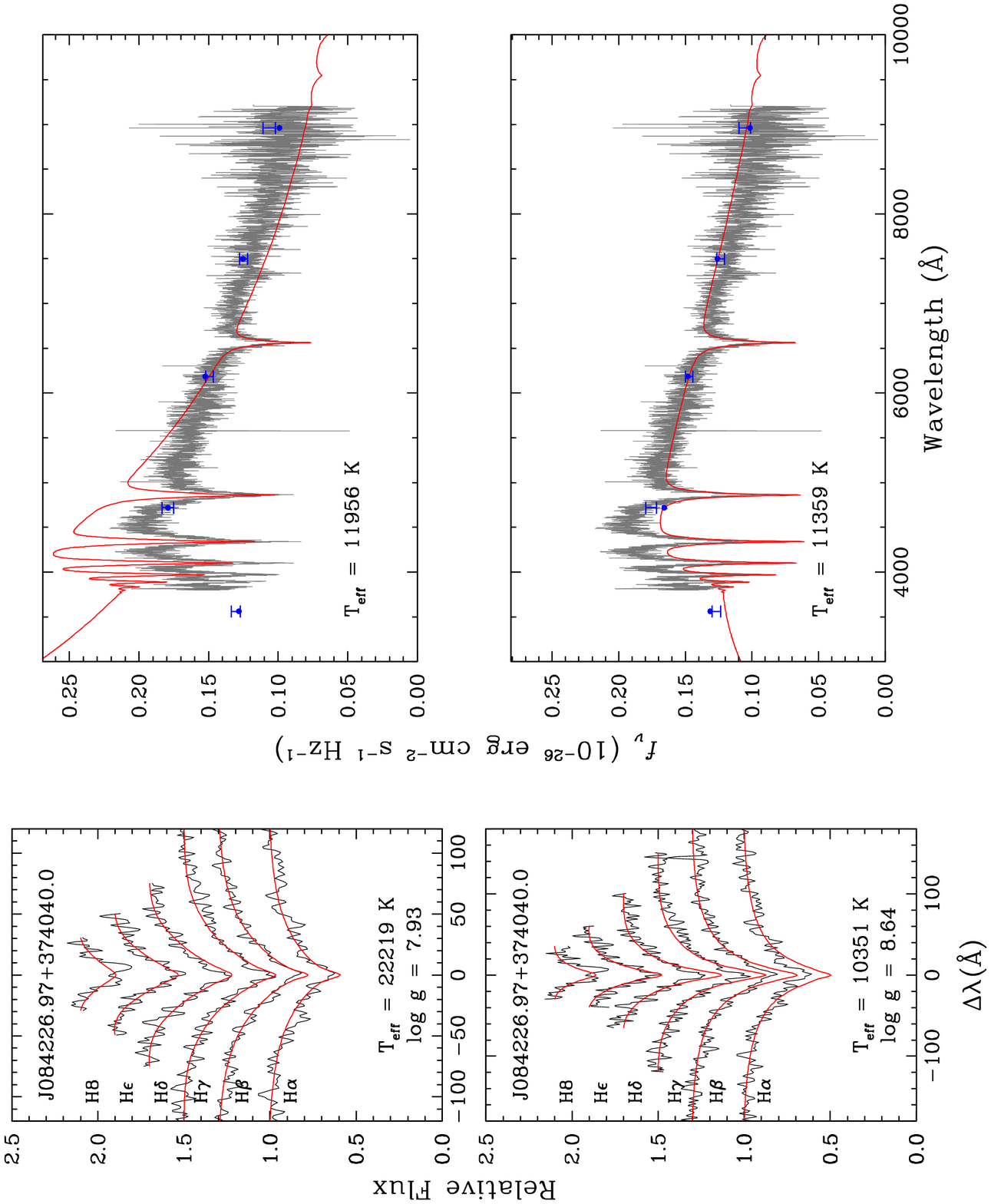] {Similar to Figure \ref{fg:f4} for both the hot
  and cool spectroscopic solutions of J084226.97+374040.0 (top and
  bottom panels, respectively). The observed and model spectra on the
  left panels are binned by a factor of two for clarity. This object
  is actually a DA+DC binary candidate.
\label{fg:f5}}

\figcaption[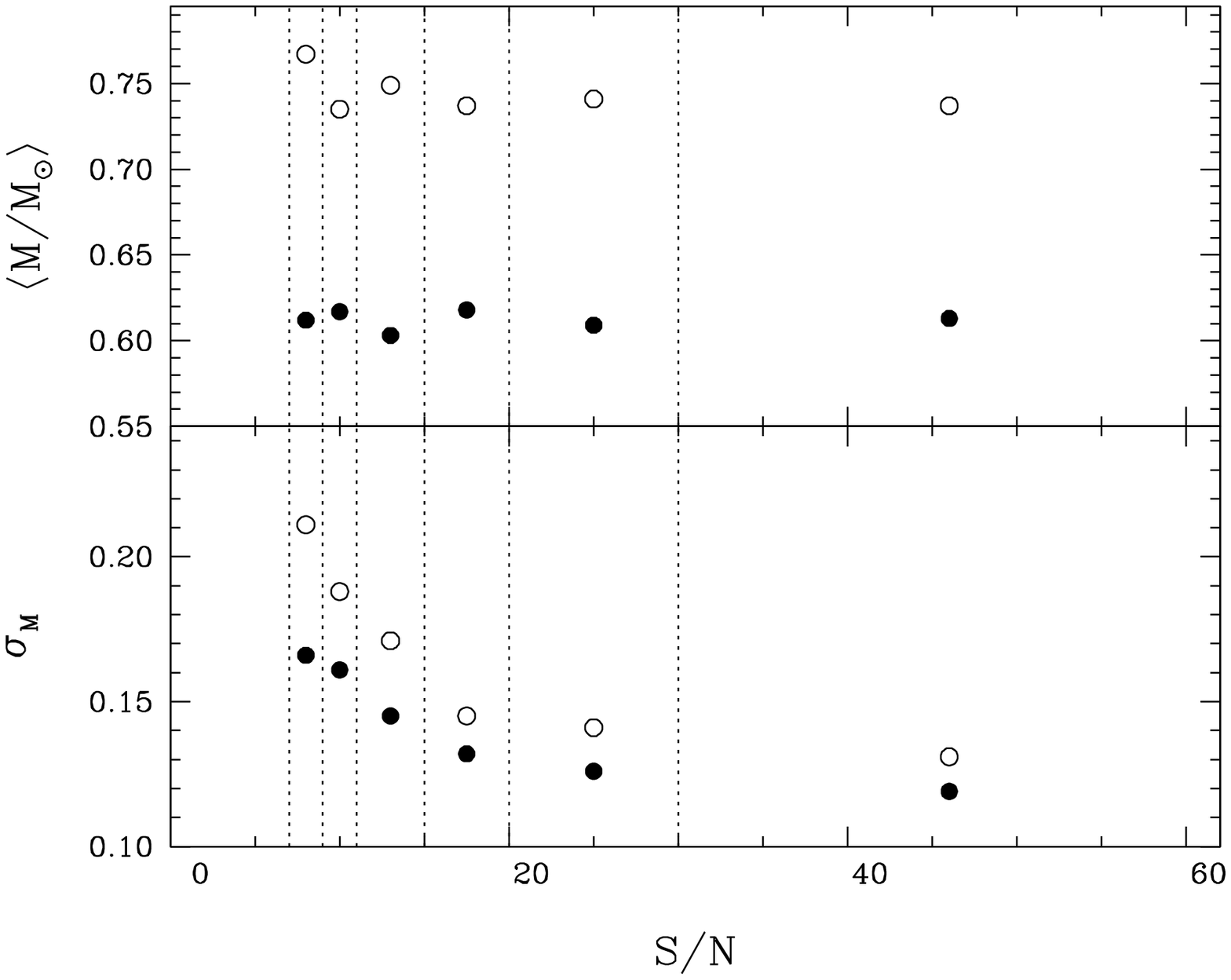] {Top panel: mean mass of the DA stars in the SDSS
 as a function of the S/N of the observations. The objects have been
 separated in bins of nearly equal number of stars, identified by
 dotted vertical lines. Filled circles correspond to DA stars with
 40,000 K $> \Te >$ 13,000 K, while open circles represent cooler
 objects. Bottom panel: similar to the top panel but for the mass
 standard deviation as a function S/N.
\label{fg:f6}}

\figcaption[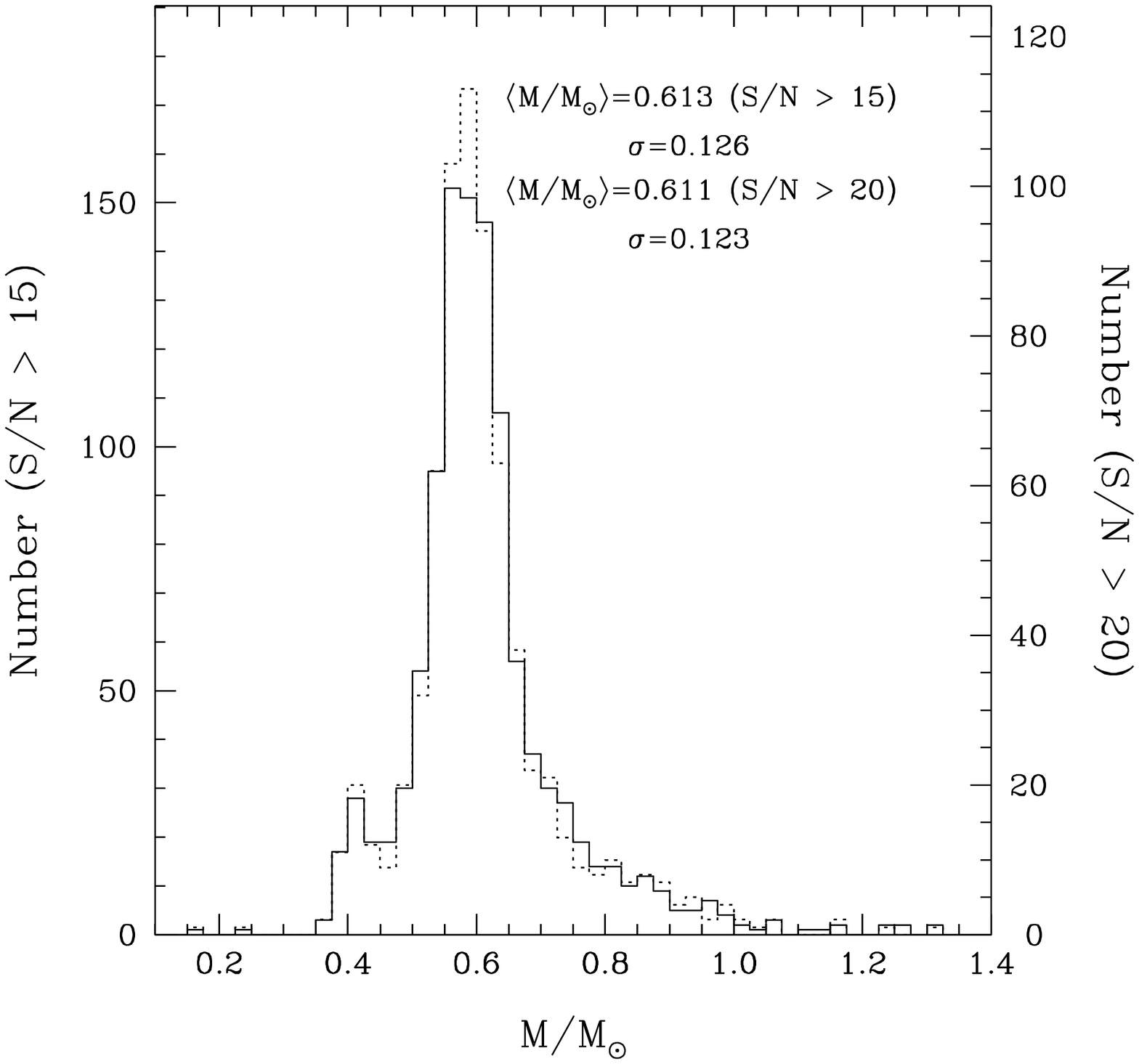] {Mass distribution for the DA stars in the SDSS
  with 40,000 K $> \Te >$ 13,000 K. The distribution shown with a
  solid line corresponds to our optimal sample of 1089 DA stars with
  ${\rm S/N} > 15$. In comparison, we show as a dashed line the
  distribution with an alternate cutoff of ${\rm S/N} > 20$, scaled to
  match the former (the number of stars is given on the right-hand
  scale). The mean mass and standard deviation are given in the
  figure.
\label{fg:f7}}

\figcaption[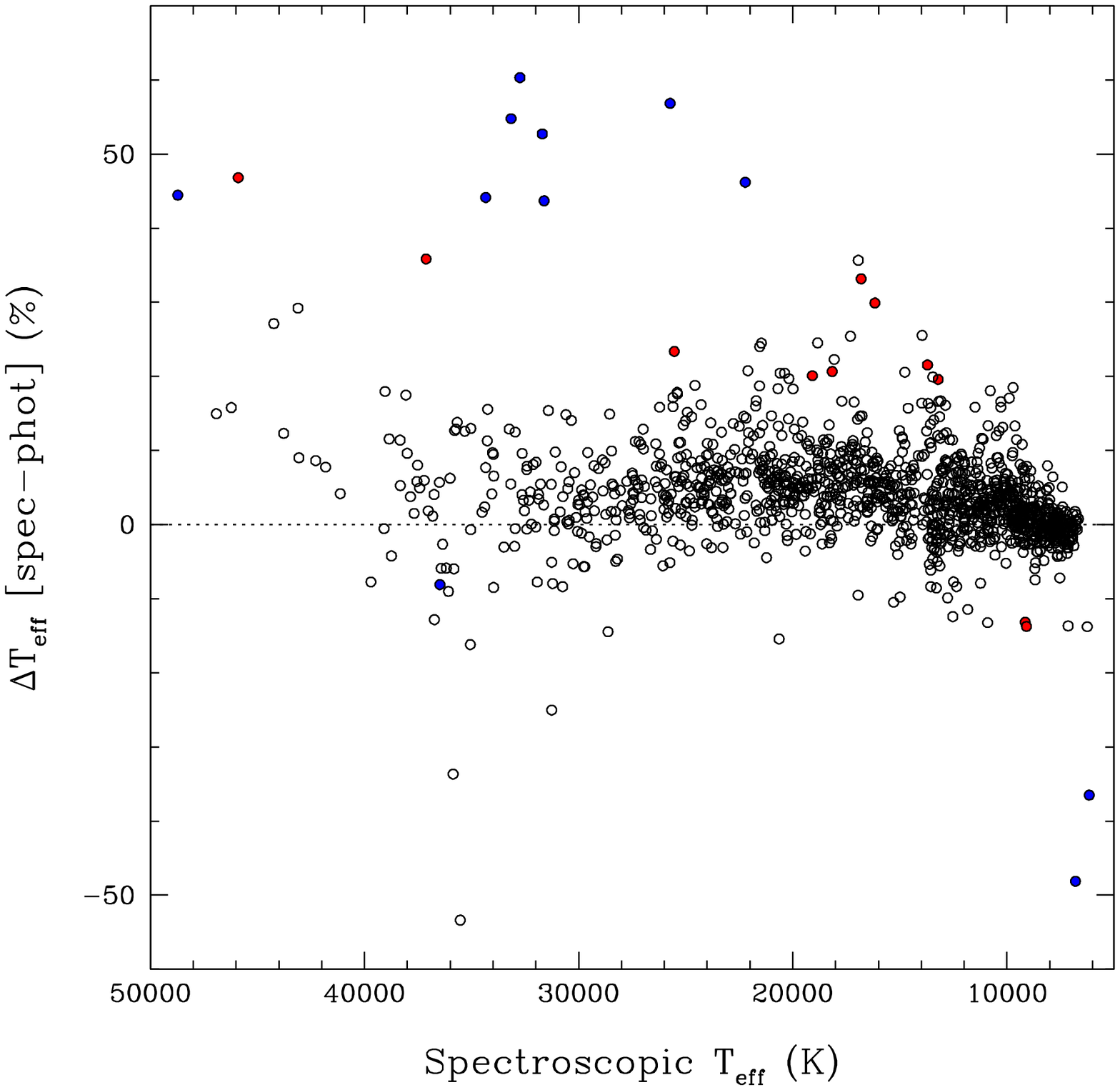] {Comparison of spectroscopic and photometric
temperatures (in $\%$ with respect to spectroscopic temperatures) as a
function of spectroscopic temperature for all DA stars with ${\rm S/N}
> 20$ (open circles), excluding DA$-$M dwarf binaries. Blue filled
circles correspond to double degenerate candidates discussed in the text,
while red symbols represent outliers (2$\sigma$ discrepancy)
that have been flagged in Table 1
as problematic observations --- see Notes (1) or (2) in Table 1. The
horizontal dotted line represents a perfect match between both
temperature estimates.
\label{fg:f8}}

\figcaption[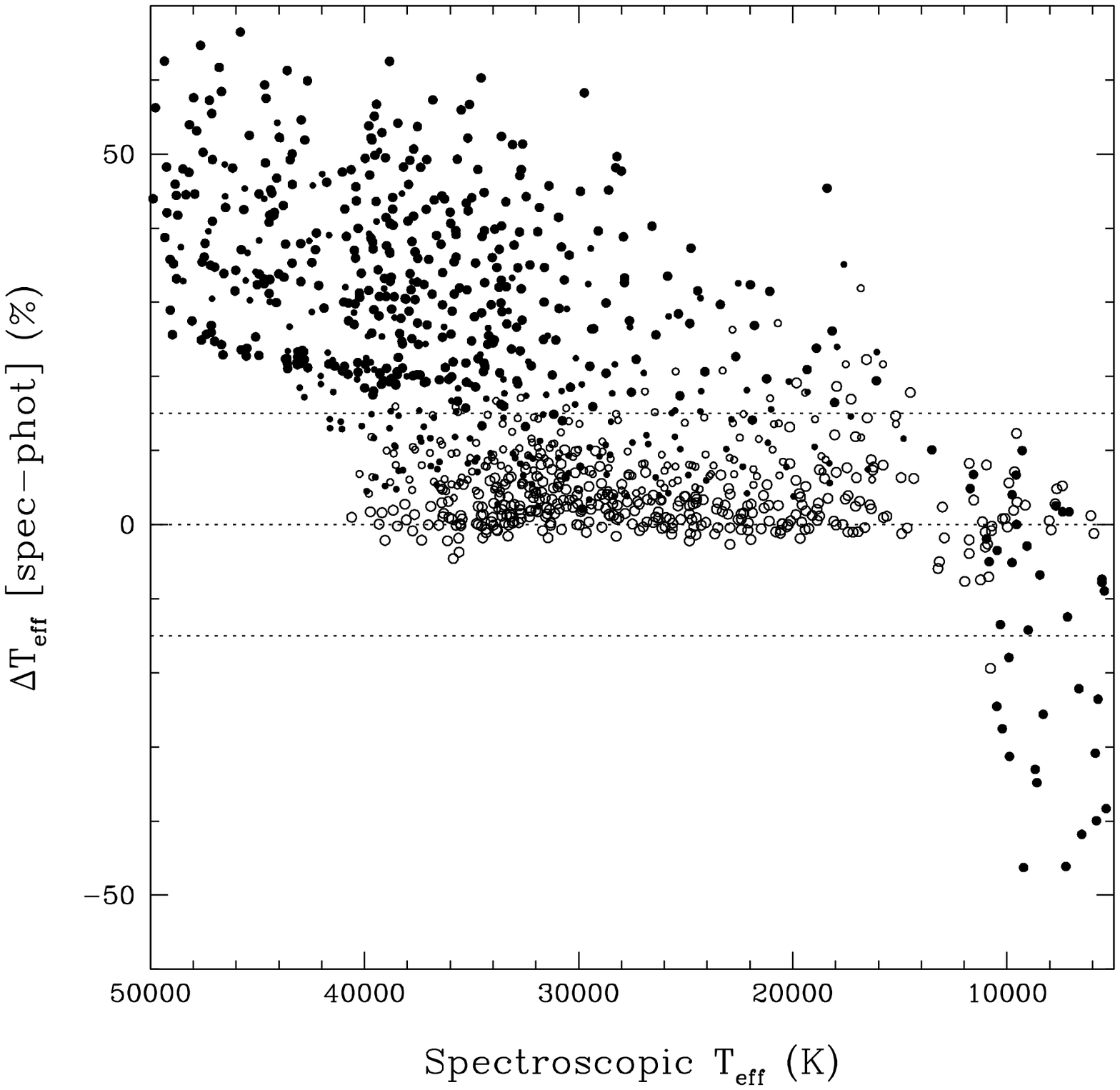] {Simulation of DA+DA (open circles) and DA+DB/DC
 (filled circles) binaries with synthetic models computed for every
 component combination in the range of 40,000 K $> \Te >$ 6000 K with
 steps of 2000 K. We studied the cases of two equal surface gravities
 ($\log g = 8$), and a difference of 0.5 dex ($\log g = 7.75,
 8.25$). The resulting model spectra, with noise added, are fitted using
 the same procedure as that used to analyze the SDSS stars in Figure
 \ref{fg:f8} (see text for more details). The smaller points
 correspond to simulations where the ratio in temperature between the
 hotter and cooler components is larger than a factor of two. The
 middle horizontal line represents the 1:1 relation, while the two
 other horizontal lines correspond to the $2\sigma$ region obtained
 from the SDSS distribution displayed in Figure \ref{fg:f8}.
\label{fg:f9}}

\figcaption[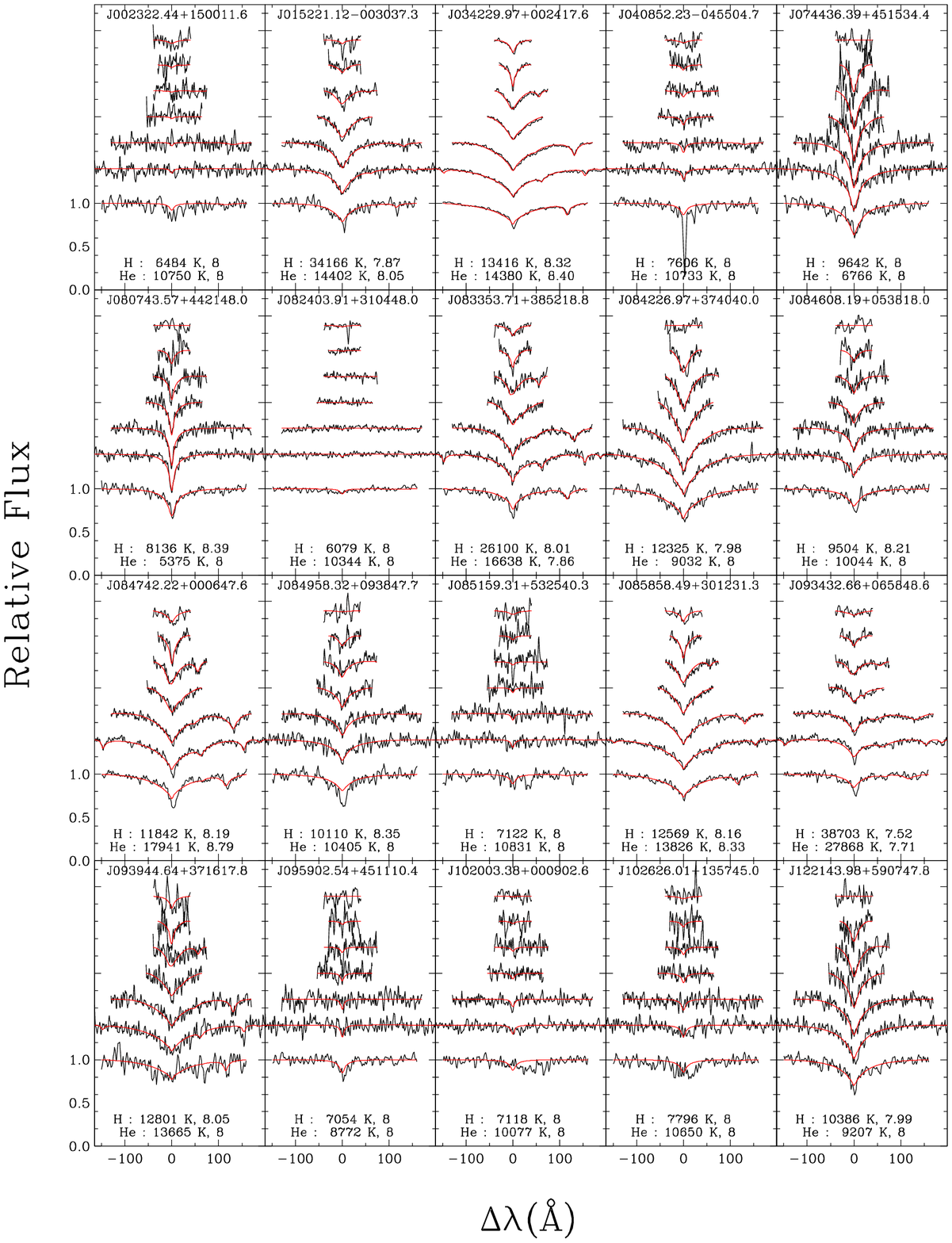] {Our best fits to the Balmer lines and He \textsc{i}
 lines, if present, for 35 double degenerate candidates in the SDSS
 sample. The lines are, from bottom to top, H$\alpha$ to H8, and He
 \textsc{i} $\lambda$5877. Both the predicted and observed spectra
 have been binned by a factor of two for clarity. The atmospheric
 parameters, $\Te$ and $\log g$, of both H-rich and He-rich components
 are given in each panel; a value of $\logg=8$ without decimals
 indicates that the value is assumed. The unexpected features in the core of
 H$\alpha$ for some objects (i.e., J040852.23$-$045504.7) are a common
 occurrence in the SDSS spectra, and the origin of these features is
 unknown to us.
\label{fg:f10}}

\figcaption[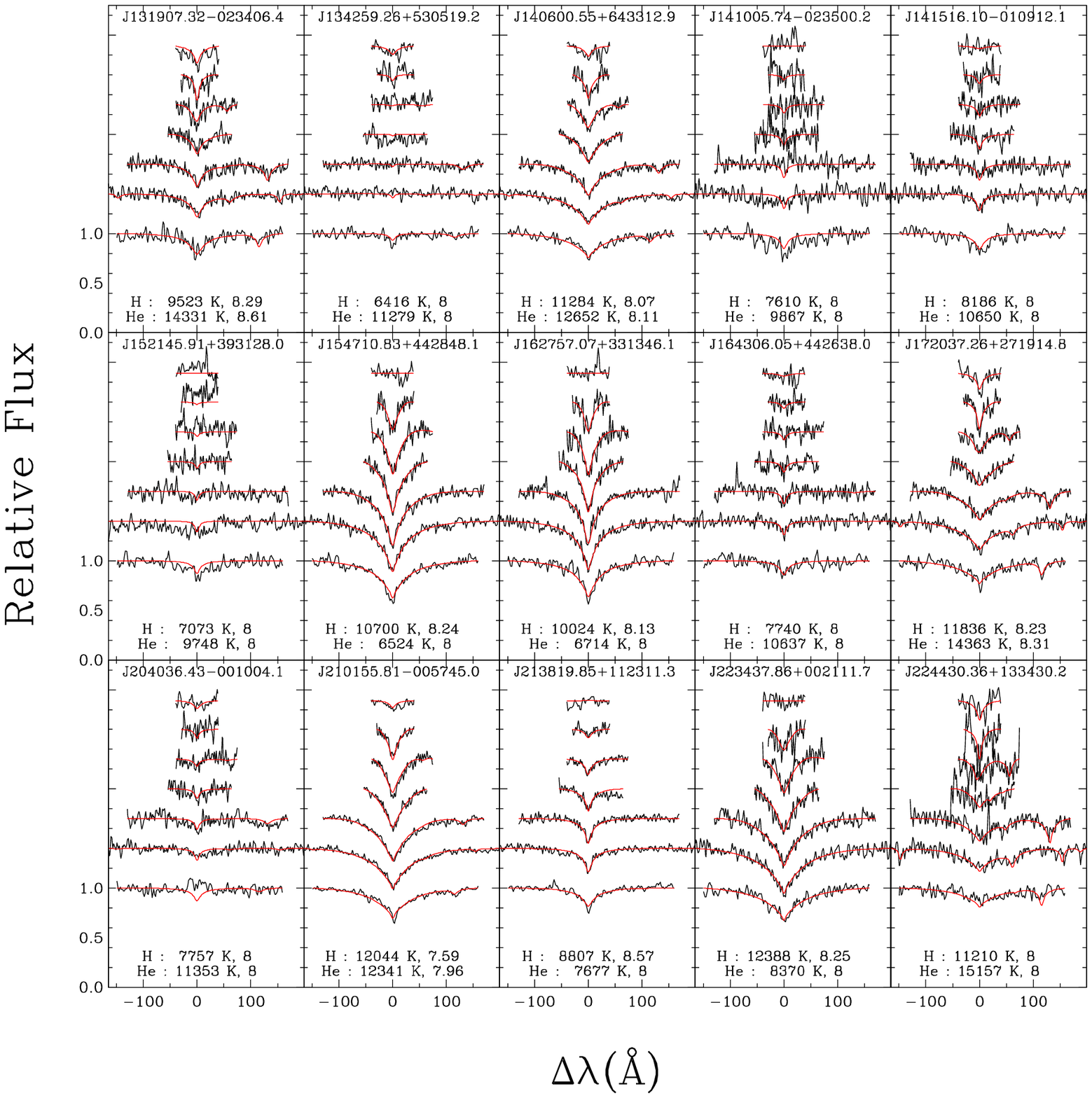] 
{Same as Figure \ref{fg:f10}.
\label{fg:f11}}

\figcaption[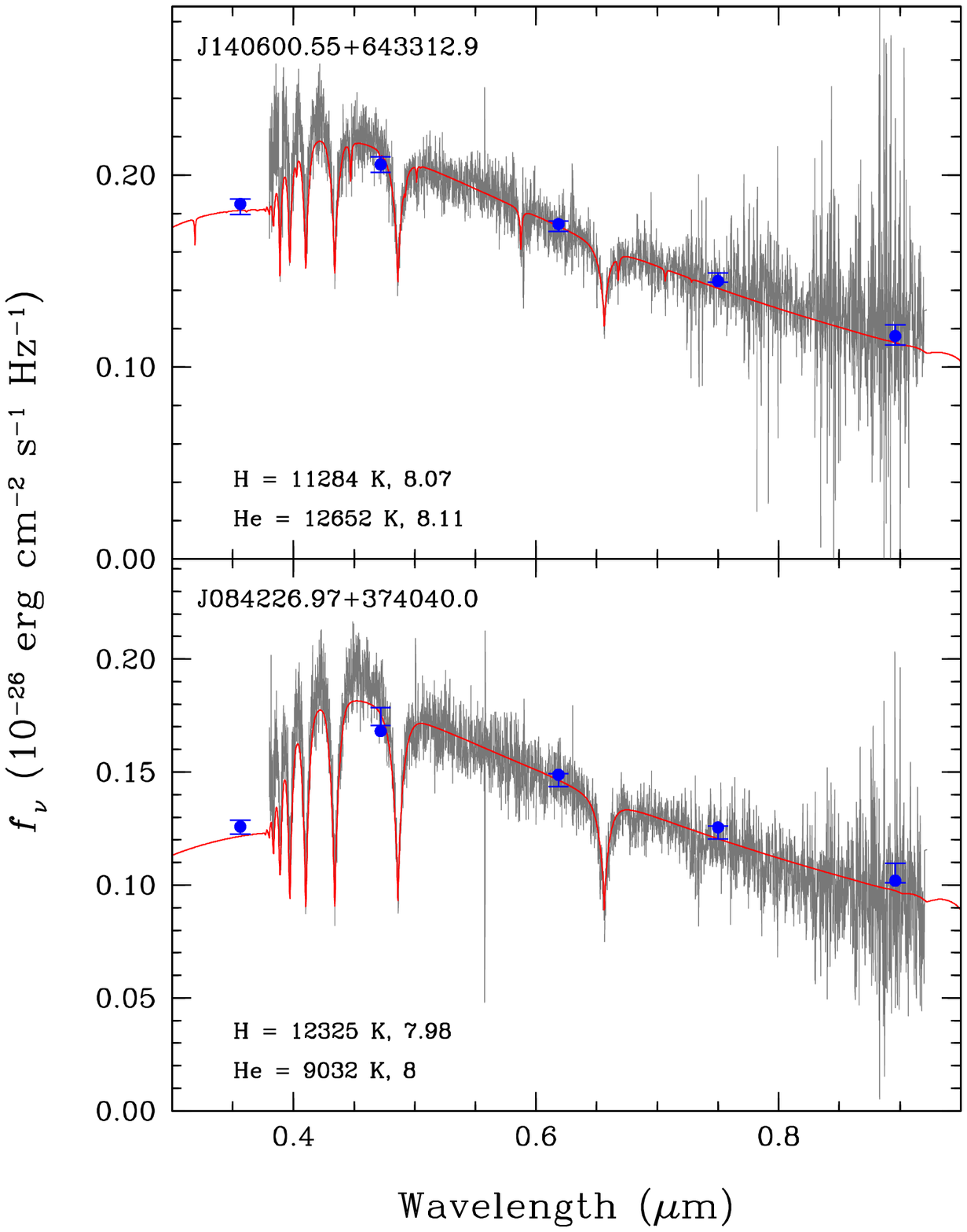] {Superposition of the predicted (blue dots) and observed
(blue error bars) photometry for two objects taken from Figures
 \ref{fg:f10} and \ref{fg:f11} (and in Table 3). The atmospheric
 parameters obtained from our spectroscopic fits are used here to
 compute the predicted photometry; only the solid angle is
 adjusted. Both the observed (gray) and synthetic (red) absolute
 fluxes are then scaled to the $r$ photometric band.
\label{fg:f12}}

\figcaption[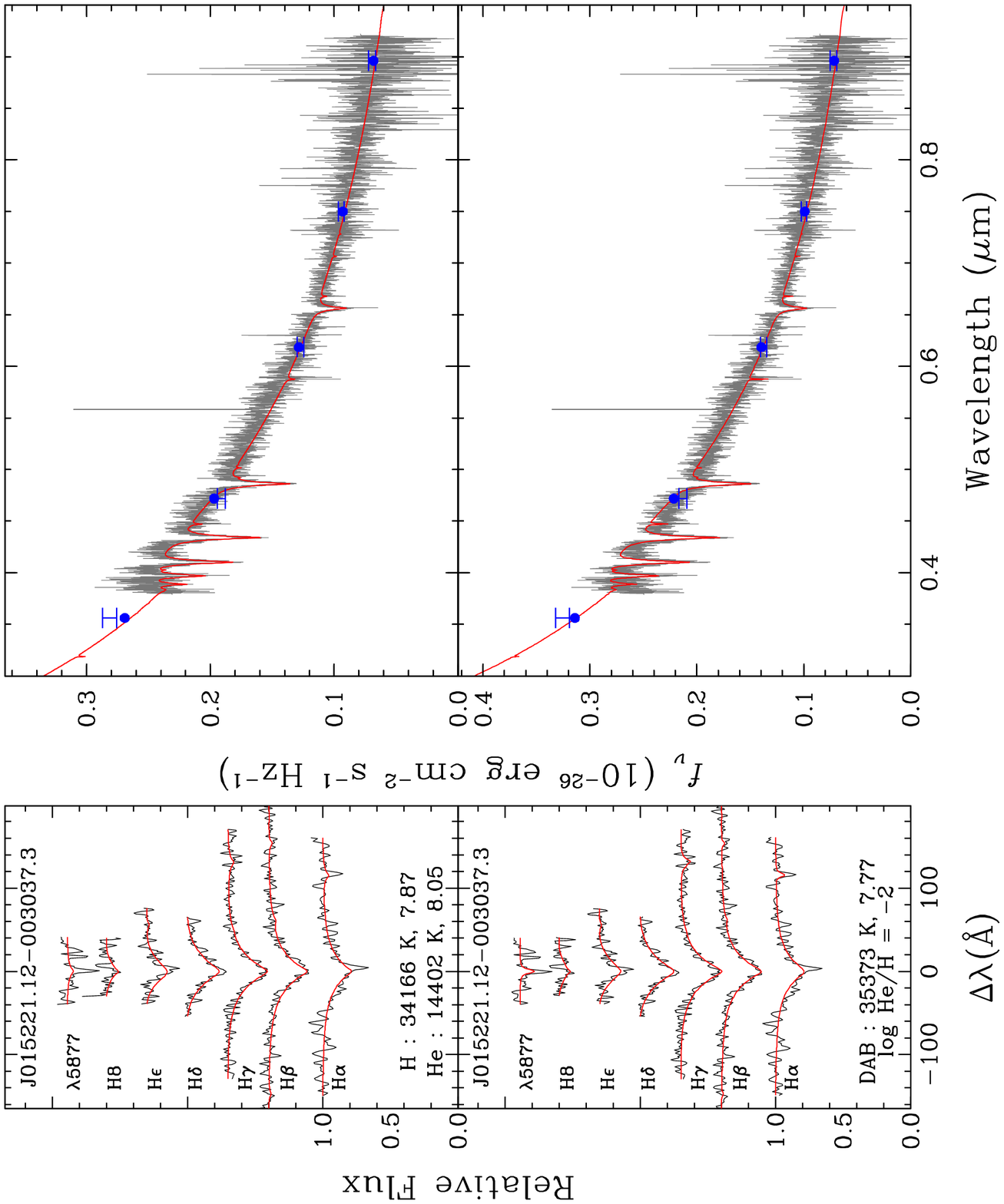] 
{Top: our DA+DB solution for J015221.12$-$003037.3. The left panel
shows our best spectroscopic fit reproduced from Figure \ref{fg:f10},
while the right panel shows the superposition of the observed and
predicted photometry (only the scaling factor is adjusted here) using
the same presentation format as previous figures. Bottom: our DAB
homogeneous solution for the same object. The left panel shows our
spectroscopic fit with an {\it assumed} helium abundance of He/H $=
10^{-2}$. Again, the right side represents a superposition of the
observed and predicted photometry using the spectroscopic atmospheric
parameters.
\label{fg:f13}}

\figcaption[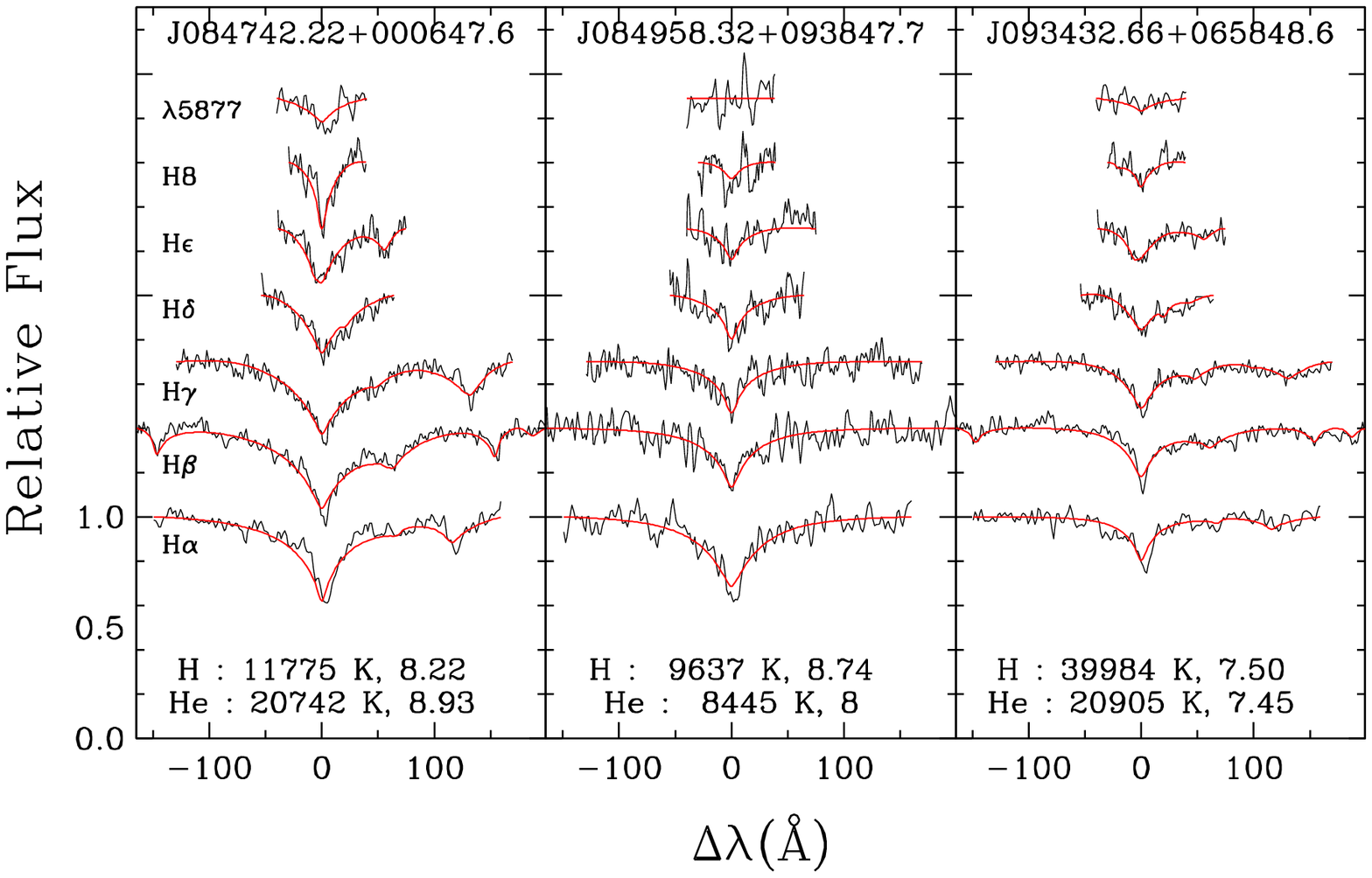] 
{Same as Figure \ref{fg:f10} for three objects but with an {\it
assumed} hydrogen abundance of H/He $= 10^{-3}$ for the helium-rich
components.
\label{fg:f14}}

\figcaption[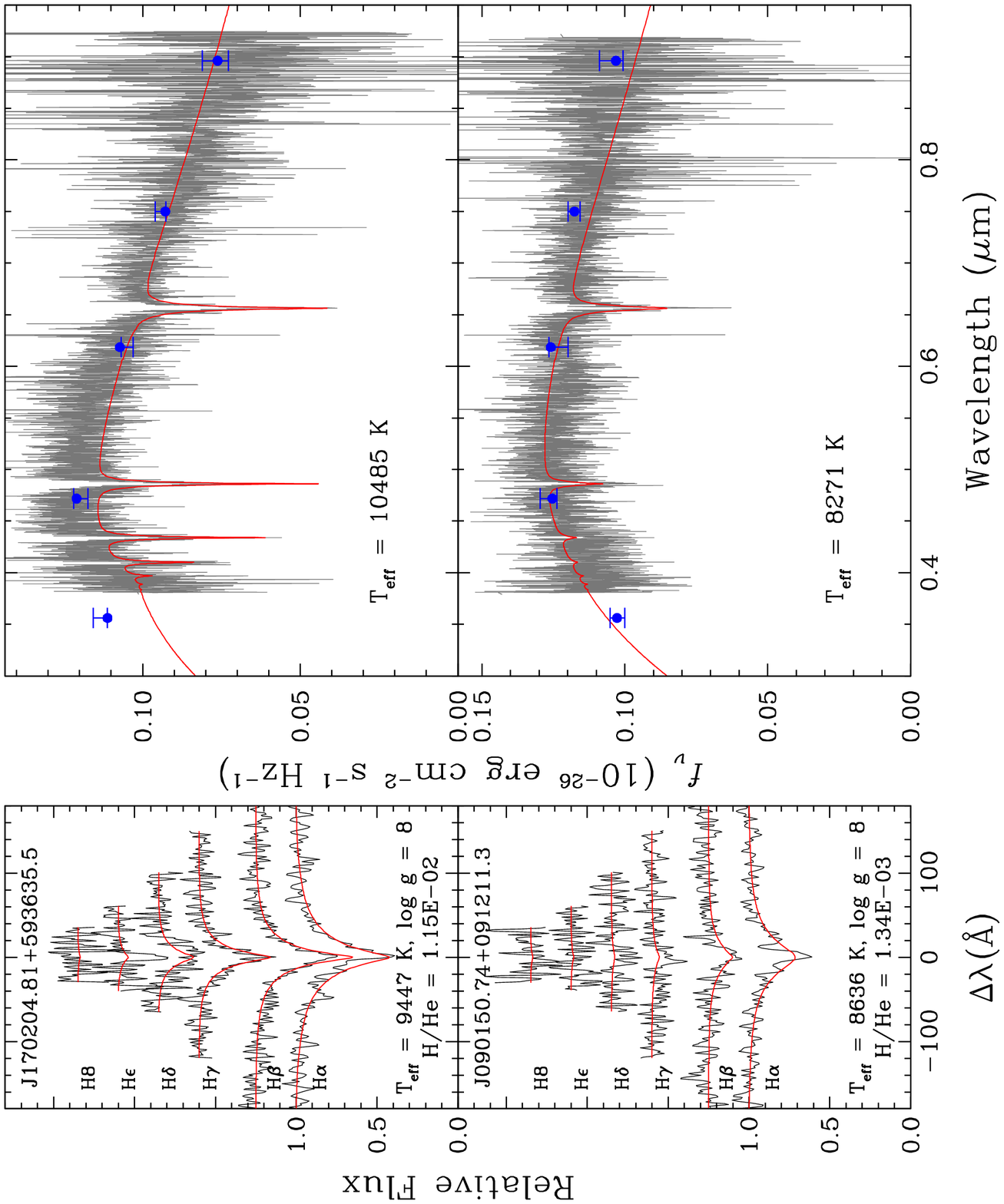] 
{Left panels: our best fits to two DA stars with mixed H/He model
atmospheres; a value of $\log g=8$ is assumed in both cases. The
atmospheric parameters are given in each panel. Both the predicted and
observed spectra have been binned by a factor of two for
clarity. Right panels: our best photometric fits for the same objects
assuming a value of $\log g = 8$ and H/He abundances determined from
the spectroscopic fits. Both the fluxed spectra (in gray) and
synthetic model fluxes (in red), calculated at the spectroscopic
parameters given in the left panel, are then scaled to the $r$
photometric band.
\label{fg:f15}}

\figcaption[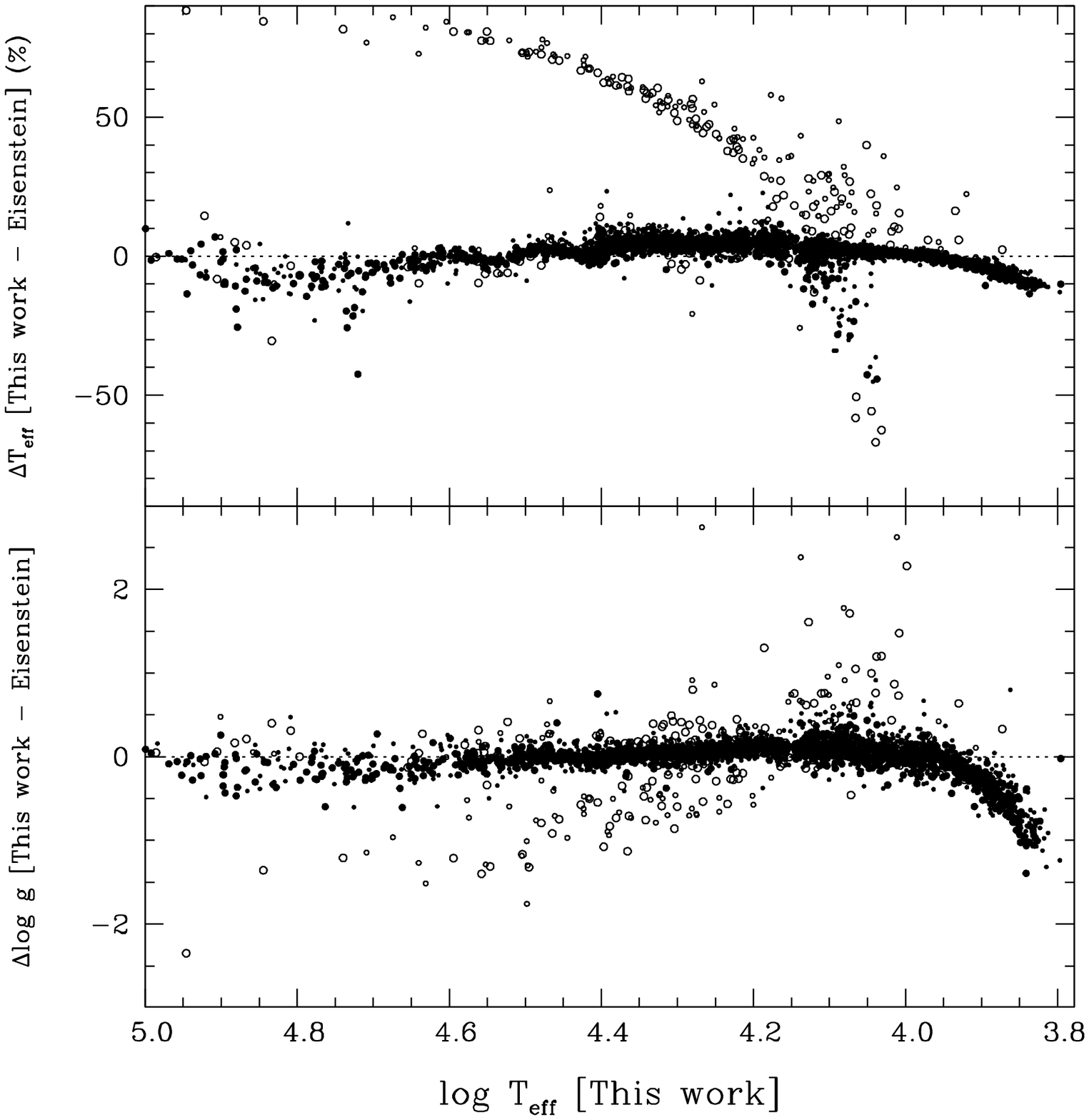] 
{Comparison of atmospheric parameters, $\Te$ and $\log g$,
between our analysis and that of \citet{E06}. The open
circles represent DA$-$M dwarf binaries, while smaller symbols
indicate spectra with $20 > {\rm S/N} > 12$. We note that the model
grid used by Eisenstein et al.~is limited to $\log g < 9$, hence some
of the discrepancies observed here result from this shortcoming.
\label{fg:f16}}

\figcaption[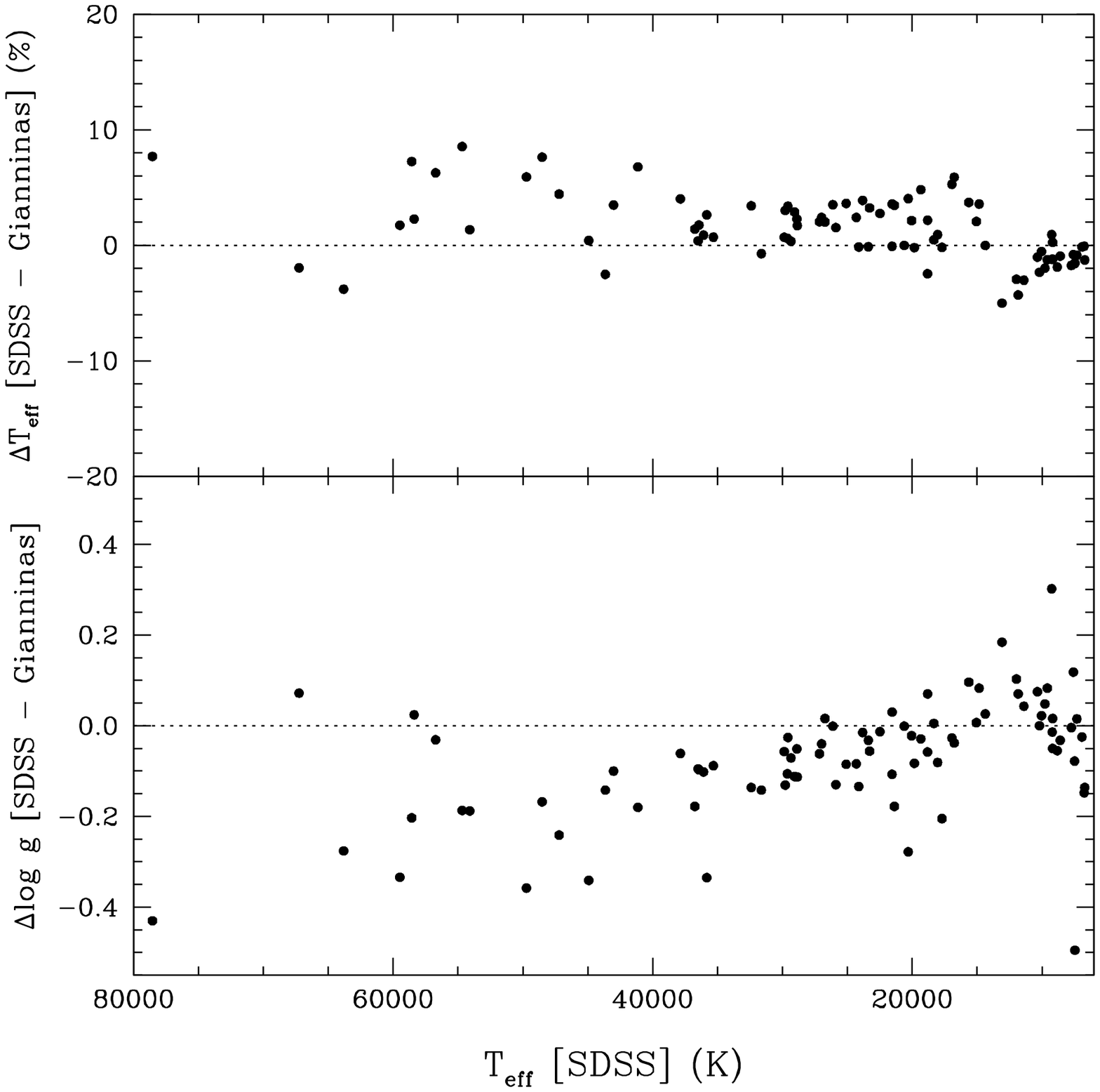] 
{Top panel: effective temperature differences (in \%) for the 89 DA stars in
common between the SDSS and Gianninas et al. samples. The observed spectra
have been analyzed with the same models and fitting technique
discussed in
Section 2. The horizontal line represents a perfect match. Bottom
panel: similar to the top panel but for differences in $\logg$ values.
\label{fg:f17}}

\figcaption[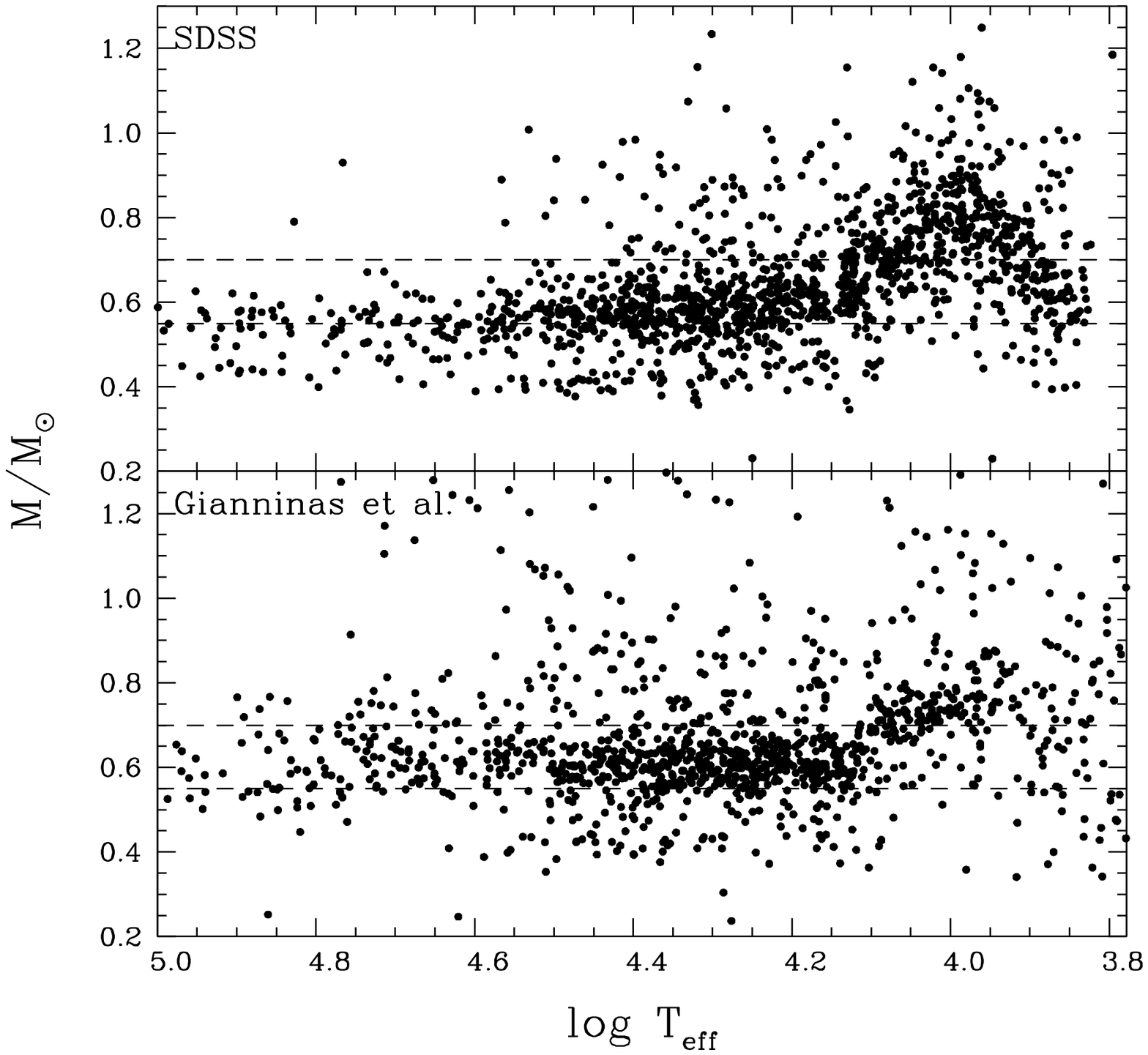] 
{Mass distributions as a function of $\Te$ for the SDSS sample
(top panel) and the Gianninas et al.~sample (bottom panel). Lines of
constant mass at 0.55 $M_{\odot}$ and 0.70 $M_{\odot}$ are shown as a
reference.
\label{fg:f18}}

\figcaption[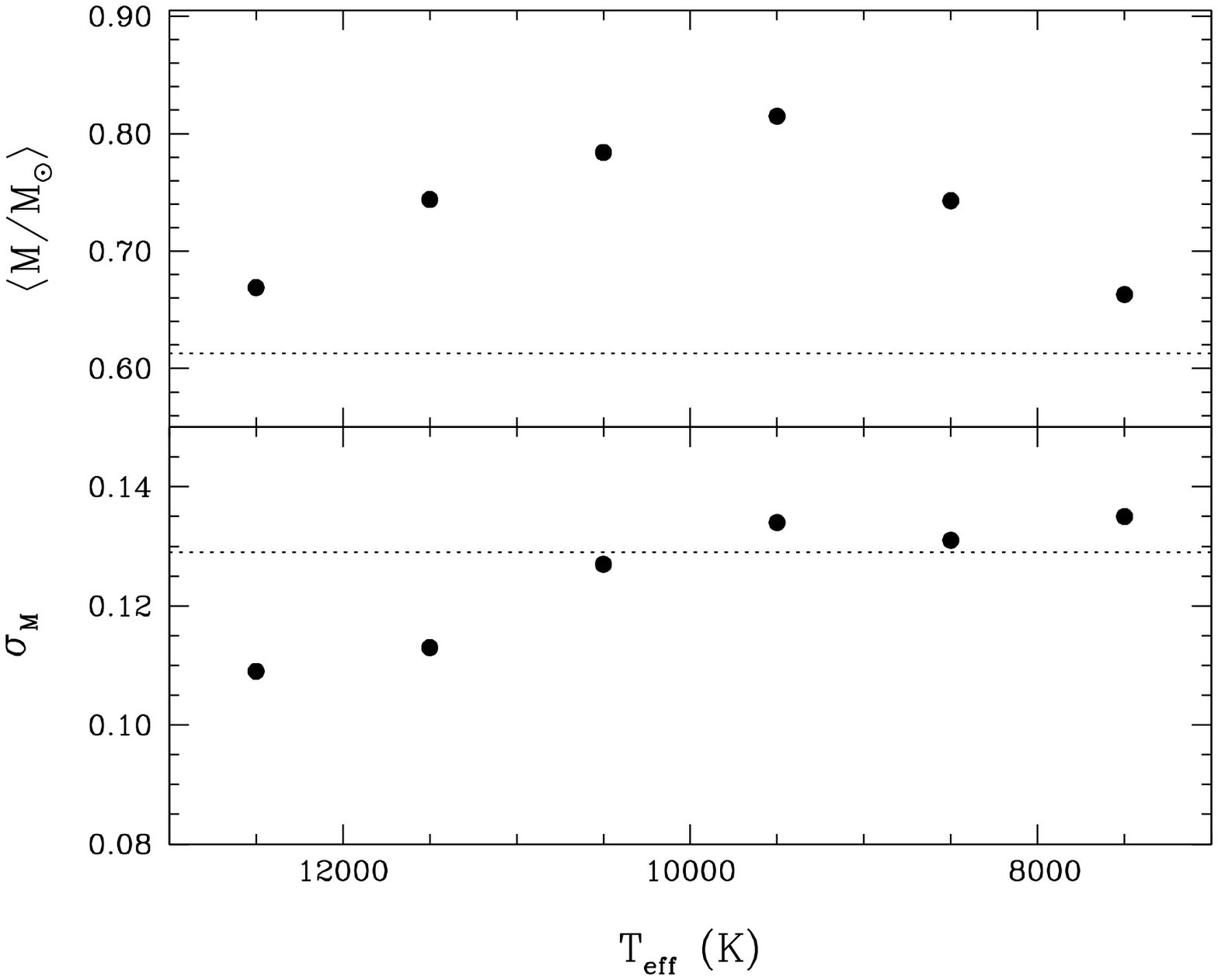] {Top panel: mean mass of the SDSS sample for
  $T_{\rm eff} < 13,000$ K computed in 1000 K temperature bins. The
  dotted line is the mean mass obtained from Figure \ref{fg:f7} for
  hotter objects. Bottom panel: similar to the top panel but for the
  mass standard deviation.
\label{fg:f19}}

\figcaption[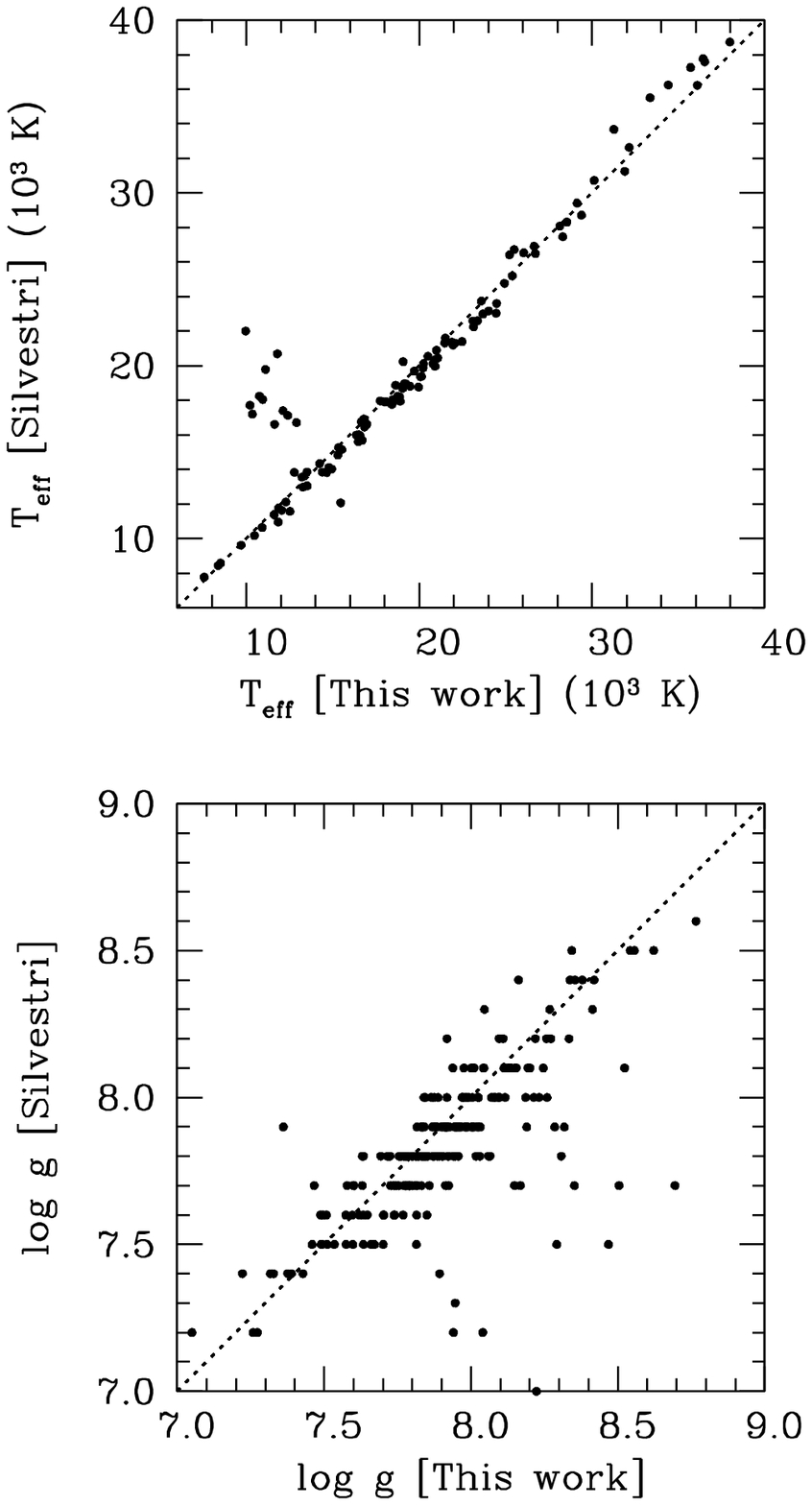] 
{Comparison of our $\Te$ and $\log g$ determinations 
with those of \citet{silvestri06} for the sample of
DA$-$M dwarf binaries in the SDSS. The dashed line
in each panel represents the 1:1 correspondence.
\label{fg:f20}}

\figcaption[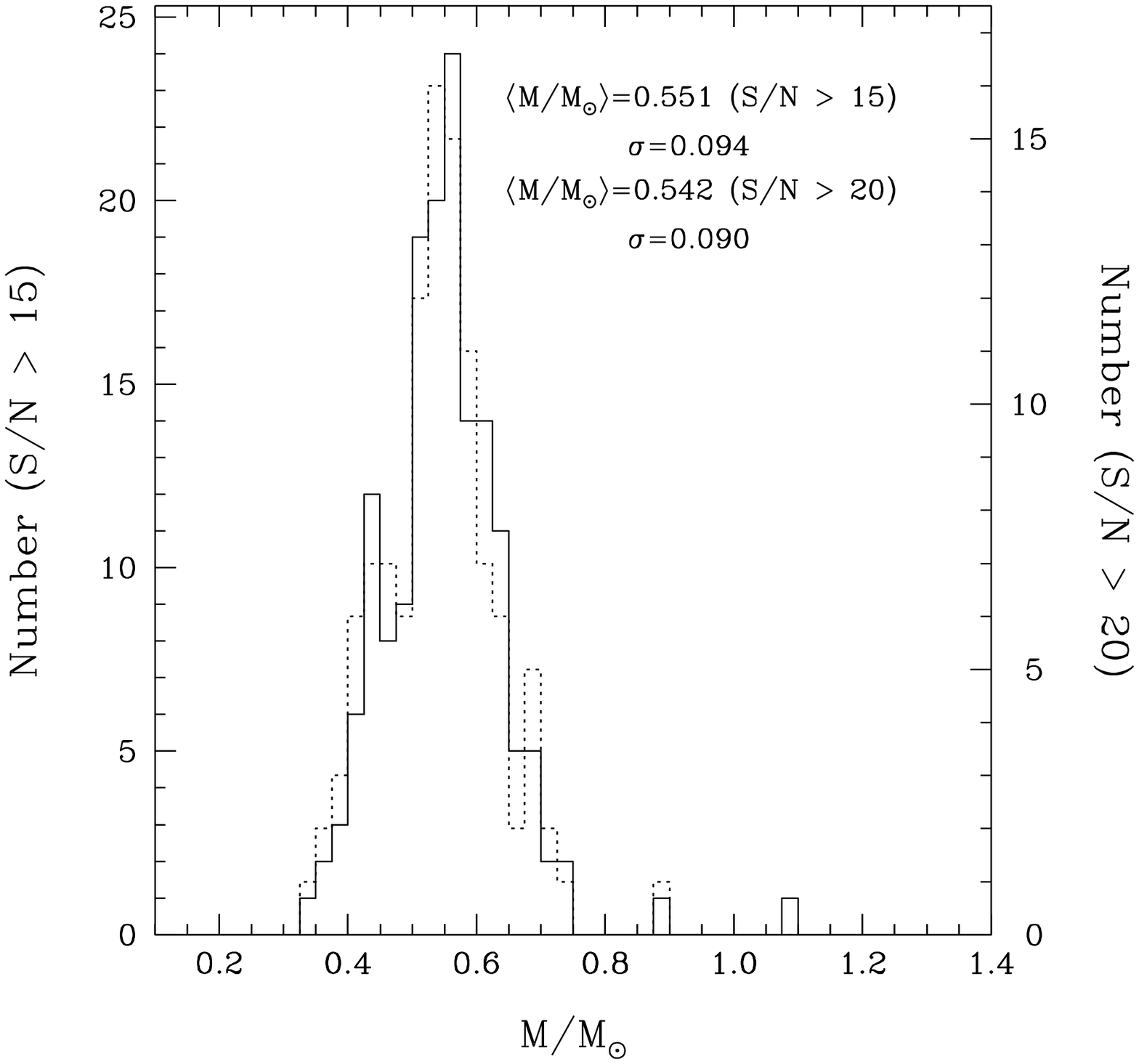] 
{Same as Figure \ref{fg:f7} but for the mass distribution of our
sample of DA$-$M dwarf binaries with a DA component in the range 40,000 K $> \Te >$
13,000 K.
\label{fg:f21}}

\begin{figure}[p]
\plotone{f1.eps}
\begin{flushright}
Figure \ref{fg:f1}
\end{flushright}
\end{figure}

\begin{figure}[p]
\plotone{f2.eps}
\begin{flushright}
Figure \ref{fg:f2}
\end{flushright}
\end{figure}

\begin{figure}[p]
\plotone{f3.eps}
\begin{flushright}
Figure \ref{fg:f3}
\end{flushright}
\end{figure}

\begin{figure}[l]
\includegraphics[angle=270,scale=0.65]{f4.eps}
\begin{flushright}
Figure \ref{fg:f4}
\end{flushright}
\end{figure}

\begin{figure}[l]
\includegraphics[angle=270,scale=0.65]{f5.eps}
\begin{flushright}
Figure \ref{fg:f5}
\end{flushright}
\end{figure}

\begin{figure}[p]
\plotone{f6.eps}
\begin{flushright}
Figure \ref{fg:f6}
\end{flushright}
\end{figure}

\begin{figure}[p]
\plotone{f7.eps}
\begin{flushright}
Figure \ref{fg:f7}
\end{flushright}
\end{figure}

\begin{figure}[p]
\plotone{f8.eps}
\begin{flushright}
Figure \ref{fg:f8}
\end{flushright}
\end{figure}

\begin{figure}[p]
\plotone{f9.eps}
\begin{flushright}
Figure \ref{fg:f9}
\end{flushright}
\end{figure}

\begin{figure}[p]
\plotone{f10.eps}
\begin{flushright}
Figure \ref{fg:f10}
\end{flushright}
\end{figure}

\begin{figure}[p]
\plotone{f11.eps}
\begin{flushright}
Figure \ref{fg:f11}
\end{flushright}
\end{figure}

\begin{figure}[p]
\plotone{f12.eps}
\begin{flushright}
Figure \ref{fg:f12}
\end{flushright}
\end{figure}

\begin{figure}[l]
\includegraphics[angle=270,scale=0.65]{f13.eps}
\begin{flushright}
Figure \ref{fg:f13}
\end{flushright}
\end{figure}

\begin{figure}[p]
\plotone{f14.eps}
\begin{flushright}
Figure \ref{fg:f14}
\end{flushright}
\end{figure}

\begin{figure}[l]
\includegraphics[angle=270,scale=0.65]{f15.eps}
\begin{flushright}
Figure \ref{fg:f15}
\end{flushright}
\end{figure}

\begin{figure}[p]
\plotone{f16.eps}
\begin{flushright}
Figure \ref{fg:f16}
\end{flushright}
\end{figure}

\begin{figure}[p]
\plotone{f17.eps}
\begin{flushright}
Figure \ref{fg:f17}
\end{flushright}
\end{figure}

\begin{figure}[p]
\plotone{f18.eps}
\begin{flushright}
Figure \ref{fg:f18}
\end{flushright}
\end{figure}

\clearpage

\begin{figure}[p]
\plotone{f19.eps}
\begin{flushright}
Figure \ref{fg:f19}
\end{flushright}
\end{figure}
\clearpage

\begin{figure}[p]
\plotone{f20.eps}
\begin{flushright}
Figure \ref{fg:f20}
\end{flushright}
\end{figure}

\begin{figure}[p]
\plotone{f21.eps}
\begin{flushright}
Figure \ref{fg:f21}
\end{flushright}
\end{figure}

\end{document}